\documentclass[aps,prd,superscriptaddress,showpacs,nofootinbib,floatfix,twocolumn]{revtex4}

\usepackage{graphicx}	

\usepackage{xspace}	

\newcommand{\pt}{\mbox{$p_T$}\xspace}
\newcommand{\mt}{\mbox{$m_T$}\xspace}

\newcommand{\mpt}{\mbox{$\langle p_T \rangle$}\xspace}
\newcommand{\mmt}{\mbox{$\langle m_T \rangle$}\xspace}
\newcommand{\sqs}{\mbox{$\sqrt{s}$}\xspace}
\newcommand{\pp}{\mbox{$p+p$}\xspace}

\newcommand{\cs}{\mbox{$d\sigma/dy$}\xspace}
\newcommand{\bwg}{\mbox{{\it BW$*$G}}\xspace}

\newcommand{\bei}{\begin{itemize}}

\newcommand{\eei}{\end{itemize}}

\begin{document}

\title{Measurement of neutral mesons in $p+p$ collisions at $\sqrt{s}$~=~200~GeV \\
       and scaling properties of hadron production}

\newcommand{\abilene}{Abilene Christian University, Abilene, Texas 79699, USA}
\newcommand{\acadsin}{Institute of Physics, Academia Sinica, Taipei 11529, Taiwan}
\newcommand{\banaras}{Department of Physics, Banaras Hindu University, Varanasi 221005, India}
\newcommand{\barc}{Bhabha Atomic Research Centre, Bombay 400 085, India}
\newcommand{\bnlcoll}{Collider-Accelerator Department, Brookhaven National Laboratory, Upton, New York 11973-5000, USA}
\newcommand{\bnlphys}{Physics Department, Brookhaven National Laboratory, Upton, New York 11973-5000, USA}
\newcommand{\caucr}{University of California - Riverside, Riverside, California 92521, USA}
\newcommand{\charlesczech}{Charles University, Ovocn\'{y} trh 5, Praha 1, 116 36, Prague, Czech Republic}
\newcommand{\ciae}{China Institute of Atomic Energy (CIAE), Beijing, People's Republic of China}
\newcommand{\cns}{Center for Nuclear Study, Graduate School of Science, University of Tokyo, 7-3-1 Hongo, Bunkyo, Tokyo 113-0033, Japan}
\newcommand{\colorado}{University of Colorado, Boulder, Colorado 80309, USA}
\newcommand{\columbia}{Columbia University, New York, New York 10027 and Nevis Laboratories, Irvington, New York 10533, USA}
\newcommand{\czechtech}{Czech Technical University, Zikova 4, 166 36 Prague 6, Czech Republic}
\newcommand{\dapnia}{Dapnia, CEA Saclay, F-91191, Gif-sur-Yvette, France}
\newcommand{\debrecen}{Debrecen University, H-4010 Debrecen, Egyetem t{\'e}r 1, Hungary}
\newcommand{\elte}{ELTE, E{\"o}tv{\"o}s Lor{\'a}nd University, H - 1117 Budapest, P{\'a}zm{\'a}ny P. s. 1/A, Hungary}
\newcommand{\fit}{Florida Institute of Technology, Melbourne, Florida 32901, USA}
\newcommand{\fsu}{Florida State University, Tallahassee, Florida 32306, USA}
\newcommand{\gsu}{Georgia State University, Atlanta, Georgia 30303, USA}
\newcommand{\hiroshima}{Hiroshima University, Kagamiyama, Higashi-Hiroshima 739-8526, Japan}
\newcommand{\ihepprot}{IHEP Protvino, State Research Center of Russian Federation, Institute for High Energy Physics, Protvino, 142281, Russia}
\newcommand{\illuiuc}{University of Illinois at Urbana-Champaign, Urbana, Illinois 61801, USA}
\newcommand{\instpasczech}{Institute of Physics, Academy of Sciences of the Czech Republic, Na Slovance 2, 182 21 Prague 8, Czech Republic}
\newcommand{\isu}{Iowa State University, Ames, Iowa 50011, USA}
\newcommand{\jinrdubna}{Joint Institute for Nuclear Research, 141980 Dubna, Moscow Region, Russia}
\newcommand{\kek}{KEK, High Energy Accelerator Research Organization, Tsukuba, Ibaraki 305-0801, Japan}
\newcommand{\kfki}{KFKI Research Institute for Particle and Nuclear Physics of the Hungarian Academy of Sciences (MTA KFKI RMKI), H-1525 Budapest 114, POBox 49, Budapest, Hungary}
\newcommand{\korea}{Korea University, Seoul, 136-701, Korea}
\newcommand{\kurchatov}{Russian Research Center ``Kurchatov Institute", Moscow, Russia}
\newcommand{\kyoto}{Kyoto University, Kyoto 606-8502, Japan}
\newcommand{\labllr}{Laboratoire Leprince-Ringuet, Ecole Polytechnique, CNRS-IN2P3, Route de Saclay, F-91128, Palaiseau, France}
\newcommand{\lawllnl}{Lawrence Livermore National Laboratory, Livermore, California 94550, USA}
\newcommand{\losalamos}{Los Alamos National Laboratory, Los Alamos, New Mexico 87545, USA}
\newcommand{\lpc}{LPC, Universit{\'e} Blaise Pascal, CNRS-IN2P3, Clermont-Fd, 63177 Aubiere Cedex, France}
\newcommand{\lund}{Department of Physics, Lund University, Box 118, SE-221 00 Lund, Sweden}
\newcommand{\mass}{Department of Physics, University of Massachusetts, Amherst, Massachusetts 01003-9337, USA }
\newcommand{\muenster}{Institut f\"ur Kernphysik, University of Muenster, D-48149 Muenster, Germany}
\newcommand{\muhlenberg}{Muhlenberg College, Allentown, Pennsylvania 18104-5586, USA}
\newcommand{\myongji}{Myongji University, Yongin, Kyonggido 449-728, Korea}
\newcommand{\nagasaki}{Nagasaki Institute of Applied Science, Nagasaki-shi, Nagasaki 851-0193, Japan}
\newcommand{\newmex}{University of New Mexico, Albuquerque, New Mexico 87131, USA }
\newcommand{\nmsu}{New Mexico State University, Las Cruces, New Mexico 88003, USA}
\newcommand{\ornl}{Oak Ridge National Laboratory, Oak Ridge, Tennessee 37831, USA}
\newcommand{\orsay}{IPN-Orsay, Universite Paris Sud, CNRS-IN2P3, BP1, F-91406, Orsay, France}
\newcommand{\peking}{Peking University, Beijing, People's Republic of China}
\newcommand{\pnpi}{PNPI, Petersburg Nuclear Physics Institute, Gatchina, Leningrad region, 188300, Russia}
\newcommand{\riken}{RIKEN Nishina Center for Accelerator-Based Science, Wako, Saitama 351-0198, JAPAN}
\newcommand{\rikjrbrc}{RIKEN BNL Research Center, Brookhaven National Laboratory, Upton, New York 11973-5000, USA}
\newcommand{\rikkyo}{Physics Department, Rikkyo University, 3-34-1 Nishi-Ikebukuro, Toshima, Tokyo 171-8501, Japan}
\newcommand{\saispbstu}{Saint Petersburg State Polytechnic University, St. Petersburg, Russia}
\newcommand{\saopaulo}{Universidade de S{\~a}o Paulo, Instituto de F\'{\i}sica, Caixa Postal 66318, S{\~a}o Paulo CEP05315-970, Brazil}
\newcommand{\seoulnat}{System Electronics Laboratory, Seoul National University, Seoul, Korea}
\newcommand{\stonybrkc}{Chemistry Department, Stony Brook University, Stony Brook, SUNY, New York 11794-3400, USA}
\newcommand{\stonycrkp}{Department of Physics and Astronomy, Stony Brook University, SUNY, Stony Brook, New York 11794, USA}
\newcommand{\subatech}{SUBATECH (Ecole des Mines de Nantes, CNRS-IN2P3, Universit{\'e} de Nantes) BP 20722 - 44307, Nantes, France}
\newcommand{\tenn}{University of Tennessee, Knoxville, Tennessee 37996, USA}
\newcommand{\titech}{Department of Physics, Tokyo Institute of Technology, Oh-okayama, Meguro, Tokyo 152-8551, Japan}
\newcommand{\tsukuba}{Institute of Physics, University of Tsukuba, Tsukuba, Ibaraki 305, Japan}
\newcommand{\vandy}{Vanderbilt University, Nashville, Tennessee 37235, USA}
\newcommand{\waseda}{Waseda University, Advanced Research Institute for Science and Engineering, 17 Kikui-cho, Shinjuku-ku, Tokyo 162-0044, Japan}
\newcommand{\weizmann}{Weizmann Institute, Rehovot 76100, Israel}
\newcommand{\yonsei}{Yonsei University, IPAP, Seoul 120-749, Korea}
\affiliation{\abilene}
\affiliation{\acadsin}
\affiliation{\banaras}
\affiliation{\barc}
\affiliation{\bnlcoll}
\affiliation{\bnlphys}
\affiliation{\caucr}
\affiliation{\charlesczech}
\affiliation{\ciae}
\affiliation{\cns}
\affiliation{\colorado}
\affiliation{\columbia}
\affiliation{\czechtech}
\affiliation{\dapnia}
\affiliation{\debrecen}
\affiliation{\elte}
\affiliation{\fit}
\affiliation{\fsu}
\affiliation{\gsu}
\affiliation{\hiroshima}
\affiliation{\ihepprot}
\affiliation{\illuiuc}
\affiliation{\instpasczech}
\affiliation{\isu}
\affiliation{\jinrdubna}
\affiliation{\kek}
\affiliation{\kfki}
\affiliation{\korea}
\affiliation{\kurchatov}
\affiliation{\kyoto}
\affiliation{\labllr}
\affiliation{\lawllnl}
\affiliation{\losalamos}
\affiliation{\lpc}
\affiliation{\lund}
\affiliation{\mass}
\affiliation{\muenster}
\affiliation{\muhlenberg}
\affiliation{\myongji}
\affiliation{\nagasaki}
\affiliation{\newmex}
\affiliation{\nmsu}
\affiliation{\ornl}
\affiliation{\orsay}
\affiliation{\peking}
\affiliation{\pnpi}
\affiliation{\riken}
\affiliation{\rikjrbrc}
\affiliation{\rikkyo}
\affiliation{\saispbstu}
\affiliation{\saopaulo}
\affiliation{\seoulnat}
\affiliation{\stonybrkc}
\affiliation{\stonycrkp}
\affiliation{\subatech}
\affiliation{\tenn}
\affiliation{\titech}
\affiliation{\tsukuba}
\affiliation{\vandy}
\affiliation{\waseda}
\affiliation{\weizmann}
\affiliation{\yonsei}
\author{A.~Adare} \affiliation{\colorado}
\author{S.~Afanasiev} \affiliation{\jinrdubna}
\author{C.~Aidala} \affiliation{\columbia} \affiliation{\mass}
\author{N.N.~Ajitanand} \affiliation{\stonybrkc}
\author{Y.~Akiba} \affiliation{\riken} \affiliation{\rikjrbrc}
\author{H.~Al-Bataineh} \affiliation{\nmsu}
\author{J.~Alexander} \affiliation{\stonybrkc}
\author{K.~Aoki} \affiliation{\kyoto} \affiliation{\riken}
\author{L.~Aphecetche} \affiliation{\subatech}
\author{R.~Armendariz} \affiliation{\nmsu}
% ...
\collaboration{PHENIX Collaboration} \noaffiliation

\date{\today}

\begin{abstract}
The PHENIX experiment at the Relativistic Heavy Ion Collider has 
measured the invariant differential cross section for production of 
$K_{S}^{0}$, $\omega$, $\eta'$, and $\phi$ mesons in $p+p$ collisions at 
$\sqrt{s}$~=~200~GeV. Measurements of $\omega$ and $\phi$ production in 
different decay channels give consistent results. New results for the 
$\omega$ are in agreement with previously published data and extend the 
measured $p_T$ coverage. The spectral shapes of all hadron transverse 
momentum distributions measured by PHENIX are well described by a 
Tsallis distribution functional form with only two parameters, $n$ and 
$T$, determining the high-$p_T$ and characterizing the low-$p_T$ regions of 
the spectra, respectively. The values of these parameters are very 
similar for all analyzed meson spectra, but with a lower parameter $T$ 
extracted for protons. The integrated invariant cross sections 
calculated from the fitted distributions are found to be consistent 
with existing measurements and with statistical model predictions. 
\end{abstract}

	

\maketitle


\section{Introduction \label{sec:intro}}

The PHENIX experiment at the Relativistic Heavy Ion Collider (RHIC) at 
Brookhaven National Laboratory has measured the production of a wide 
variety of hadrons ($\pi$, $K$, $\eta$, $\eta'$, $\omega$, $\phi$, $p$, 
$J/\psi$, and $\psi'$) at midrapidity in \pp collisions at \sqs = 200~GeV. 
The measurements were performed using a time-of-flight technique for 
charged hadron identification and via reconstruction of various photonic, 
hadronic, and dielectron decay modes for neutral hadrons. The measured 
transverse momentum spectra extend over the range from zero to 20~GeV/$c$. 
Precise measurements of hadron production in \pp collisions are crucial 
for a deeper understanding of QCD phenomena such as parton dynamics and 
hadronization. They also provide a valuable baseline for particle and jet 
production in heavy ion collisions, essential to the needs of the RHIC 
heavy ion program.

There exists a large body of experimental data on hadron production in \pp 
collisions measured at the ISR, Sp$\bar{{\rm p}}$S, Tevatron, and 
RHIC~\cite{ppg030,ppg055,ppg063,ppg064,ppg069,ppg115,star_high_pt,star_long,star_strange,star_strange_b,star_p_k_pi,tevatron-1,tevatron-2,isr-1,isr-2,isr-3,isr-4,Guettler:1976fc,Acosta:2005pk}. 
At high \pt the spectra display a power law behavior that becomes more and 
more evident as the interaction energy increases. In this regime, the 
spectra are well described by perturbative QCD together with measured 
proton structure functions~\cite{RevModPhys.67.157}. At low \pt, typically 
$\pt<2$~GeV/$c$, a region which accounts for the bulk of the produced 
particles, the spectra are governed by processes that belong to the 
non-perturbative regime of QCD and are not yet fully understood. In this 
\pt region, the spectra reveal an exponential behavior which can be 
explained with the assumption that secondary particles are emitted from a 
thermalized system with at most short-range correlations and obeying 
Boltzmann-Gibbs statistics~\cite{Hagedorn:1965st}. In this approach, the 
inverse slope parameter $T$ can be interpreted as the temperature of the 
system. However, that would require some mechanism of local thermal 
equilibrium in \pp collisions which is not yet established. It is also 
known that the particle spectra are best described by an exponential in 
\mt rather than in \pt~\cite{Hagedorn:1983wk}. According to the 
observation that the temperature parameter $T$ is the same for different 
particles a spectral shape is also the same, the so-called \mt-scaling 
observation~\cite{Guettler:1976fc, Bartke:1976zj}.

The two regimes described here, and the \pt-region where their 
contributions are predominant are commonly designated as ``soft" and 
``hard". There is no obvious boundary between them and the question as to 
what extent the mechanisms in each region are distinct is difficult to 
address. In this paper it is shown that the spectral shapes of all hadrons 
produced in \pp collisions at \sqs=200~GeV measured by PHENIX are well 
described by one single distribution without making a distinct division 
into two regions. The Tsallis~\cite{tsallis_0} distribution, also referred 
to as Levy distribution~\cite{wilk_about_q,star_high_pt}, has only two 
parameters, $T$ and $n$, that characterize the low- and high-\pt regions 
of the spectra, respectively. This distribution has been shown by Tsallis 
to result from a postulated generalization of the Boltzmann-Gibbs entropy. 
It has been suggested to be relevant for various types of systems, such as 
those with long-range correlations, or non-ergodic filling of the 
available phase space. Boltzmann statistics and exponential distributions 
are recovered in the limit that correlations disappear. The parameter $T$ 
then recovers the usual interpretation as the temperature of the system.

In a number recent publications the Tsallis statistical distribution was 
successfully applied to describe the Heavy Ion and High Energy data in 
over a wide range of incident energies and 
centralities~\cite{Biyajima:2004ub,wilk_t,Tang:2008ud,tsallis_new,Collaboration:2010xs}. 
Physical mechanisms responsible for the successful application of the 
non-extensive statistical approach to the description of the particle 
spectra in this systems is a topic of 
discussion~\cite{PhysRevE.67.036114,PhysRevE.69.038101,PhysRevE.69.038102,Parvan:2006cr,Biro:2008fv,0295-5075-65-5-606}. 
The analysis presented in this paper uses the Tsallis formalism as a 
parameterization to describe the particle spectra and compares it with 
other parameterizations used for the spectra approximation. Common 
features and differences revealed in such an approach are data driven and 
should contribute a better understanding of particle production 
mechanisms.

The successful description of the particle spectra with the Tsallis 
distribution allows to accurately calculate the integrated particle yield 
and mean momentum, even for species measured only in a limited momentum 
range. The integrated particle abundances provide important information on 
the bulk properties of the soft particle production. In particular, the 
comparison of the particle yields to statistical model predictions can be 
used to infer the degree of chemical equilibration. In the case of heavy 
ion collisions, the success of statistical model fits to the particle 
yields~\cite{Cleymans:1992zc, BraunMunzinger:1999qy} suggests that 
chemical equilibration is essentially complete. These models have also 
been used to describe particle production in \pp 
collisions~\cite{Becattini:1997rv,we}.

In this paper we present new PHENIX results on the production of neutral 
mesons in \pp collision at \sqs=200~GeV and compare the PHENIX data with 
the parameterizations commonly used to describe particle spectra in 
relativistic \pp collisions, including the Tsallis parametrization. It is 
demonstrated that the latter one describes the data in the entire range of 
measured \pt most accurately. The parameter values extracted from the fits 
are given for all measured particles.

The paper is organized as follows: Section~\ref{sec:detector} gives a 
description of the PHENIX experimental setup and detector subsystems.
Section~\ref{sec:ana_part} describes the analysis methods used to measure 
the transverse momentum spectra of different hadrons for \pp collisions at 
\sqs=200~GeV.  In Section~\ref{sec:ana_scaling} the properties of the 
measured transverse momentum spectra are analyzed.  In 
Section~\ref{sec:results} the scaling properties of the particle spectra 
are discussed and the calculated integrated yields are compared with 
published results and with statistical model calculations.  The measured 
invariant cross sections are tabulated in tables given in the Appendix.


\section{PHENIX detector \label{sec:detector}}

\begin{figure*}[htb]
\includegraphics[width=0.7\linewidth]{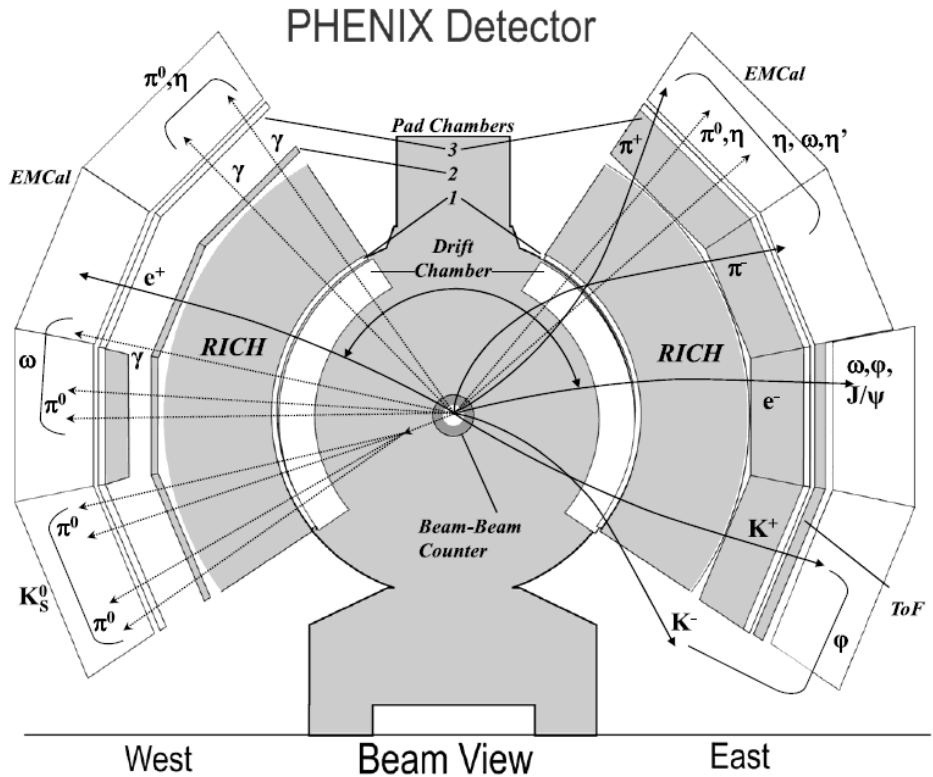}
\caption{Schematic view of the PHENIX central spectrometers and particle decay modes analyzed in this paper.\label{fig:setup}}
\end{figure*}

The PHENIX detector is designed as a high rate and fine granularity 
apparatus that utilizes a variety of detector technologies to measure 
global characteristics of the events, and to measure leptons, hadrons, and 
photons over a wide range of transverse momenta. The experimental setup 
consists of two central arm spectrometers each covering $\Delta\phi = 
\pi/2$ in azimuth at midrapidity $|\eta| < 0.35$; two forward muon 
spectrometers with full azimuthal coverage in the rapidity interval 
$1.2<|\eta|<2.4(2.2)$ for the North (South) arm and a system of "global" 
detectors. Each spectrometer provides very good momentum and spatial 
resolution and particle identification capabilities. The detailed 
description of the detector can be found elsewhere~\cite{phenix}. The 
experimental results presented in this paper were obtained using the 
central spectrometers and global detectors of the PHENIX experiment 
schematically shown in Fig.~\ref{fig:setup}.

Reconstruction of charged particle tracks and momentum measurements are 
performed with the drift chambers (DC) and the first layer of the pad 
chambers (PC1).  The fiducial volume of the DC is located outside of the 
analyzing magnetic field of the detector and has an inner radius of 2.02~m 
with an outer radius of 2.46~m. Multiple layers of wires measure the track 
position with an angular resolution of $\sim$ 0.8~mrad in the bending 
plane perpendicular to the beam axis.  The PC1, located just outside the 
outer radius of the DC, has a spatial resolution of $\sigma_{\phi}\sim$ 
2.4~mm and $\sigma_{z}\sim$ 1.7~mm and provides the $z$-coordinate of the 
track at the exit of the DC.  The momentum of a particle is determined by 
the measured bending angle in the axial magnetic field of the central 
magnet~\cite{phenix-magnet} assuming that the particle originates from the 
collision vertex. The DC momentum resolution is estimated to be $\delta 
p/p=0.7 \oplus 1.1 \% p$~[GeV/$c$]. Track matching with hits in the second 
(PC2) and third (PC3) pad chamber layers located at radii of 4.2~m and 
5.0~m, respectively, rejects tracks from secondaries originating either 
from decays of long-lived hadrons, or from interactions with the structure 
of the detector. Such tracks have not passed through the full magnetic 
field and therefore have improperly determined momenta that is typically 
overestimated.  A detailed description of the PHENIX tracking system can 
be found in~\cite{phenix-tracking,phenix-pad}.

The primary purpose of the PHENIX Electromagnetic Calorimeter (EMCal) is 
to measure the position and energy of photons and electrons. The EMCal 
covers the full acceptance of the central spectrometers and is divided 
into eight sectors. Six of the EMCal sectors located at the radius of 
5.0~m are built of lead-scintillator (PbSc) and consist of 15552 
individual towers with a granularity of 5.5$\times$5.5~cm$^{2}$ and a 
depth of 18~$X_{0}$. The two other sectors located at the radius of 5.2~m 
are built of lead-glass (PbGl) and consist of 9216 lead-glass \v{C}erenkov 
towers with a granularity of 4$\times$4~cm$^{2}$ and a depth of 
14.4~$X_{0}$. Due the fine segmentation of the EMCal the electromagnetic 
showers typically spread over several towers.  This spread provides the 
means to analyze the position and shape of the shower, and to reject 
hadrons which produce showers of a different shape.  The spatial 
resolution of the PbSc(PbGl) EMCal sector is $\sigma (E)=1.55(0.2) \oplus 
5.7(8.4) /\sqrt{E[{\rm GeV}]}$~mm for particles at normal incidence. The 
energy resolution of the PbSc(PbGl) calorimeter is $\delta
  E / E = 2.1(0.8)\% \oplus 8.1(5.9)/\sqrt{E[{\rm GeV}]}\%$.

The Time-of-Flight (TOF) subsystem is used for hadron identification based 
on momentum measurements in the DC and PC1 combined with flight path 
length from the collision vertex~\cite{phenix-id}. The TOF is located 
between the PC3 and the PbGl at the radius of 5.0~m and covers about 1/3 
of the acceptance of one central arm. The TOF detector consists of 10 
panels each containing 96 segments equipped with plastic scintillators and 
photomultiplier readout from both ends. The time resolution of $\sim$ 
120~ps enables $\pi/K$ and $K/p$ separation in the transverse momentum 
range 0.3--2.5~GeV/$c$ and 0.3--5.0~GeV/$c$, respectively. The lower limit 
is defined by the energy loss of different particles in the detector 
material.

The Ring-Imaging \v{C}erenkov (RICH) is the primary detector for $e/\pi$ 
separation. It provides an $e/\pi$ rejection factor of $\sim 10^{-3}$ for 
tracks with momenta below the pion \v{C}erenkov threshold of $\sim$ 
4~GeV/$c$ in the CO$_{2}$ used as a radiator gas.  The RICH detector in 
each arm has a mirror measuring 20~m$^{2}$ that focuses the light onto an 
array of 2560 photomultipliers. The material of the PHENIX central arm 
that precedes the RICH has been kept to just $\sim$ 2$\%$ of a radiation 
length in order to minimize the background contribution of electrons from 
$\gamma$-conversion.  The PHENIX RICH and TOF detectors are described in 
more detail in~\cite{phenix-id}.

The Beam-Beam Counters (BBC) are used for triggering, determination of the 
collision time, and location of the vertex along the beam axis, $z_{\rm 
vtx}$~\cite{phenix-bbc}. The BBC consists of two sets of 64 \v{C}erenkov 
counters surrounding the beam pipe, and located at a distance of $\pm$ 
1.44~m from the center of the interaction region. Each counter covers the 
full azimuth and the pseudorapidity interval $3.1<|\eta|<3.9$. The 
$z$-coordinate of the collision vertex is determined with a typical 
resolution of 2~cm in \pp collisions by the timing difference of the 
signals measured by the two sets of BBC counters. The time average of all 
BBC counters gives a start time for the time-of-flight measurements. The 
minimum bias trigger in \pp collisions is generated when at least one 
counter fires in each BBC set of counters, and the collision vertex 
calculated on-line is $|z_{\rm vtx}|<$ 38 cm. The efficiency of the 
minimum bias trigger is estimated to be (55$\pm$5)\% of the total 
inelastic cross section of $\sigma_{\mathit{inel}}^{pp} = 42\pm3$~mb. 
Further details about the BBC subsystem of the PHENIX detector can be 
found in~\cite{phenix-bbc}.

Due to the high rate of \pp collisions at RHIC PHENIX employs several 
specialized triggers which enable the experiment to sample more of the 
delivered luminosity for rare events. Besides the minimum bias trigger, 
the experimental results presented in this paper were obtained using the 
EMCal-RICH Trigger (ERT).

The EMCal is used to trigger on rare events with large energy deposit 
originating primarily from high energy photons or electrons.  The analog 
sum of signals from 4$\times$4 adjacent towers is compared with a trigger 
threshold of 1.4~GeV. In addition, a combination of the EMCal and the RICH 
signals is used to build the ERT trigger which is designed to select 
events containing electrons. The trigger fires when the analog sum of 
signals from 2$\times$2 adjacent towers in the EMCal exceeds a threshold 
of 0.4~GeV (setting used in the 2005 physics run) or 0.6~GeV (used in 
2006) in geometrical coincidence with a signal in the associated RICH 
trigger tile (4$\times$5 PMTs) determined using a look-up table.


\section{Neutral meson measurements \label{sec:ana_part}}

In this section we describe the analysis details of the $K_{S}^{0} 
\rightarrow \pi^{0}\pi^{0}$, $\omega \rightarrow \pi^{0}\pi^{+}\pi^{-}$, 
$\omega \rightarrow \pi^{0}\gamma$, $\omega \rightarrow e^{+}e^{-}$, 
$\eta' \rightarrow \eta\pi^{+}\pi^{-}$, $\phi \rightarrow K^{+}K^{-}$, and 
$\phi \rightarrow e^{+}e^{-}$ measurements in \pp collisions at 
\sqs=200~GeV. These measurements complete and extend previous neutral 
meson spectra results measured by the PHENIX experiment and published in 
~\cite{phenix-pi0,ppg055,ppg063,ppg069,ppg097,ppg115}

The measurements are based on a data sample representing a total 
integrated luminosity of 2.5~pb$^{-1}$ within a vertex cut of $|z_{\rm 
vtx}|<30$~cm accumulated by the PHENIX experiment in 2005.  The data were 
collected using minimum bias and ERT triggers.

\subsection{Reconstruction of neutral mesons}

Here we discuss the analysis details and main parameters of the invariant 
mass distributions reconstructed for different decay modes.

\subsubsection{Selection of the $\pi^0, \eta \rightarrow \gamma\gamma$ candidates}

\begin{figure}[htb]
\includegraphics[width=1.0\linewidth]{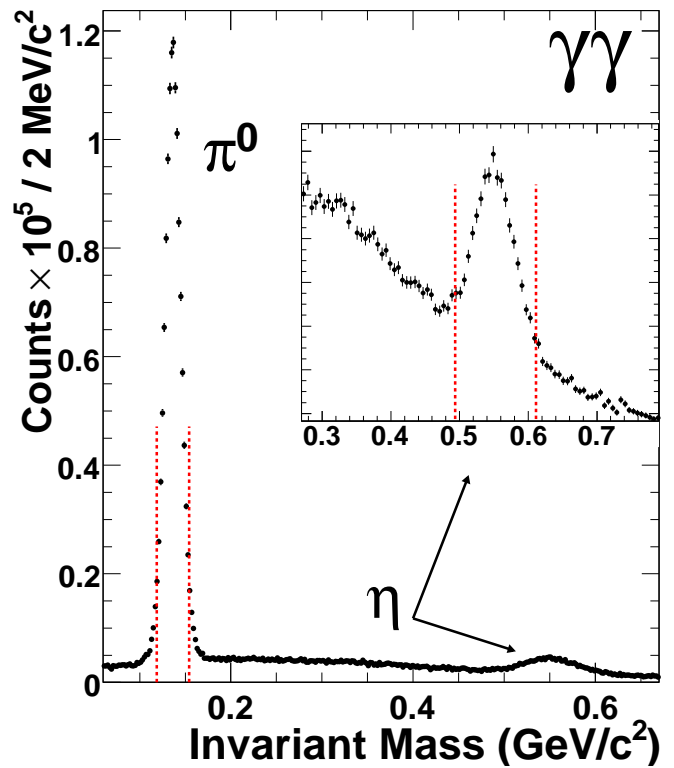}
\caption{Invariant mass distribution for $\gamma\gamma$ pairs in the
  range $4<p_{T}$~(GeV/$c)<6$.  The inset shows an enlargement of the
  region around the $\eta$ mass.}
\label{fig:pi_eta_gg}
\end{figure}

Most particles studied in this section decay producing a $\pi^{0}$
or $\eta$ meson in the final state, which in turn decays into a
$\gamma\gamma$ pair at the point of primary decay.  The analysis
procedures for the measurement of the inclusive $\pi^{0}$ and $\eta$
invariant transverse momentum spectra in \pp collisions have been
published previously~\cite{phenix-pi0,ppg063,ppg055,ppg115}. Meson
candidates were reconstructed from pairs of clusters in the EMCal with
energy $E_{\gamma}>0.2$~GeV assuming that they correspond to photons
originating from the collision vertex. A shower profile cut was used to
reject broader showers predominantly produced by
hadrons~\cite{phenix-emcal}.  The invariant mass distribution for
cluster pairs is shown in Fig.~\ref{fig:pi_eta_gg}.

The width of the peaks is determined largely by the EMCal energy 
resolution.  For $\pi^{0}(\eta)$ meson candidates the width decreases from 
12(40)~MeV/$c^{2}$ to 9(30)~MeV/$c^{2}$ between 1~GeV/$c$ and 3~GeV/$c$ of 
the pair transverse momentum.

The reconstructed positions and widths of the peaks are in agreement with 
simulation results once detector resolution and trigger biases have been 
taken into account.  The measured mass peaks were parameterized as a 
function of the $\gamma\gamma$ pair \pt. For further analyses involving 
$\pi^{0}$ or $\eta$ mesons in the final state we selected pairs with 
$p_{T}>1$~GeV/$c$ and an invariant mass within two standard deviations of 
the measured peak position. All $\gamma\gamma$ pairs satisfying these 
criteria were assigned the nominal mass of the meson~\cite{pdg} and the 
photon energies were rescaled by the ratio of the nominal to the 
reconstructed masses.

\subsubsection{$\omega \rightarrow \pi^{0}\gamma$ and $K_{S}^{0} \rightarrow \pi^{0}\pi^{0}$}

The reconstruction of $\omega \rightarrow \pi^{0}\gamma$ and $K_{S} 
\rightarrow \pi^{0}\pi^{0}$ decays was performed by combining $\pi^{0}$ 
candidates with either all other photons with energy 
$E{\gamma}>1$~GeV~\cite{ppg064} or with all other $\pi^{0}$ candidates 
from the same event. Combinations using the same EMCal clusters more than 
once were rejected.

\begin{figure*}[thb]
\includegraphics[width=0.45\linewidth]{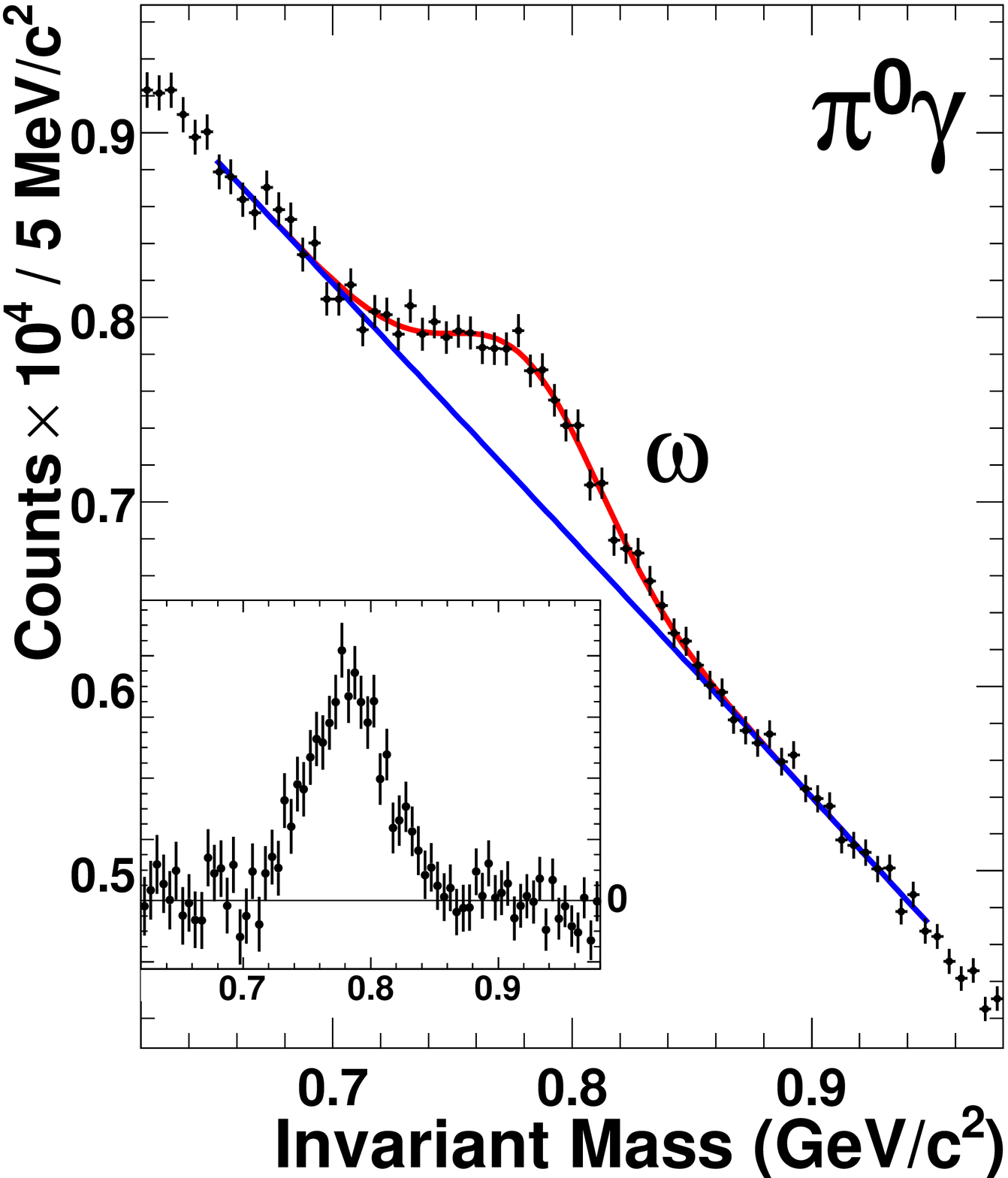} 
\includegraphics[width=0.45\linewidth]{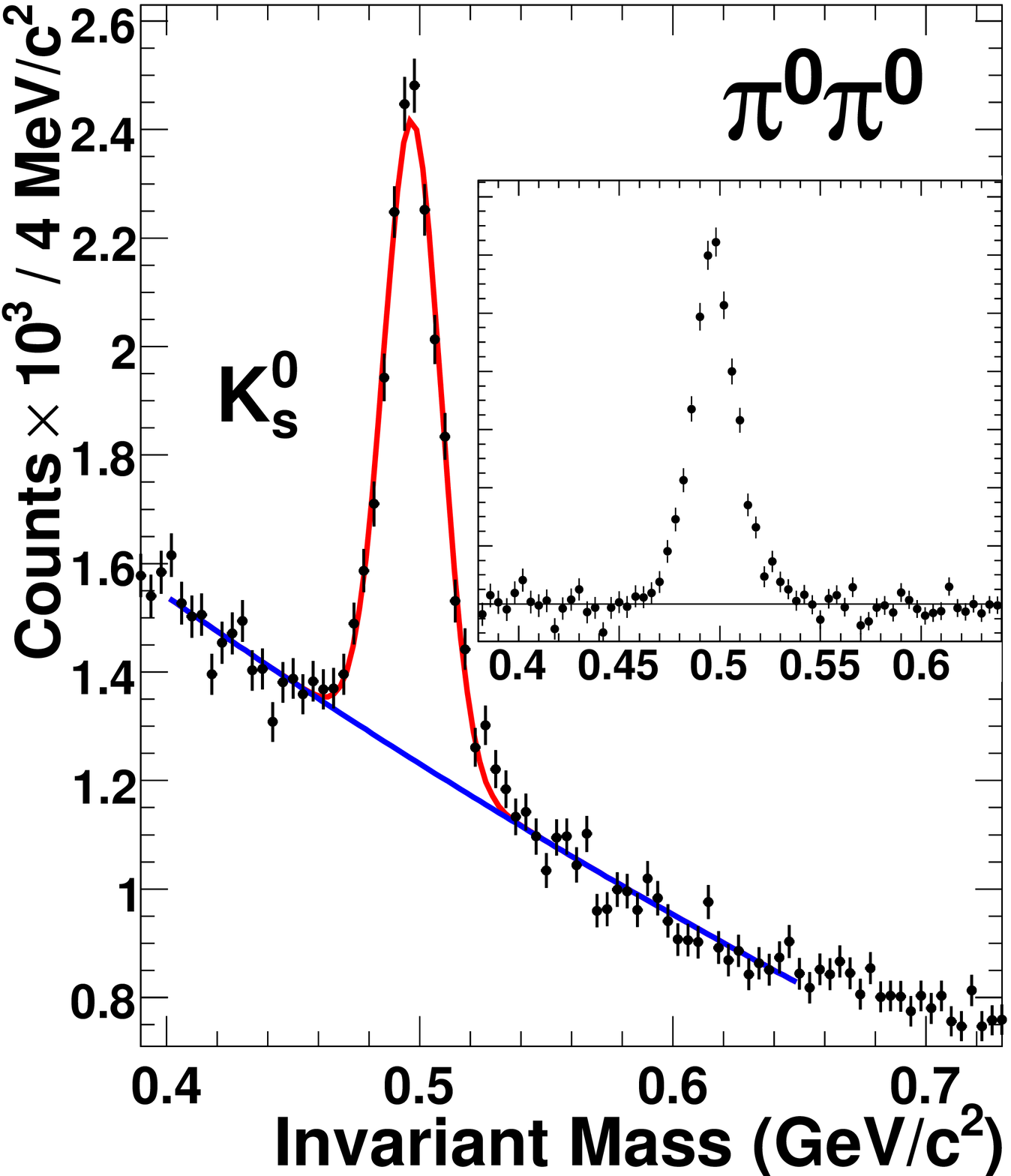}
\caption{Invariant mass distribution for $\pi^{0}\gamma$ (left) and 
$\pi^{0}\pi^{0}$ (right) decays at $4<p_{T}$~(GeV/$c)<6$.
\label{fig:minv_pi0g_pi0pi0}}
\end{figure*}

\begin{figure*}[hbt]
\includegraphics[width=0.45\linewidth]{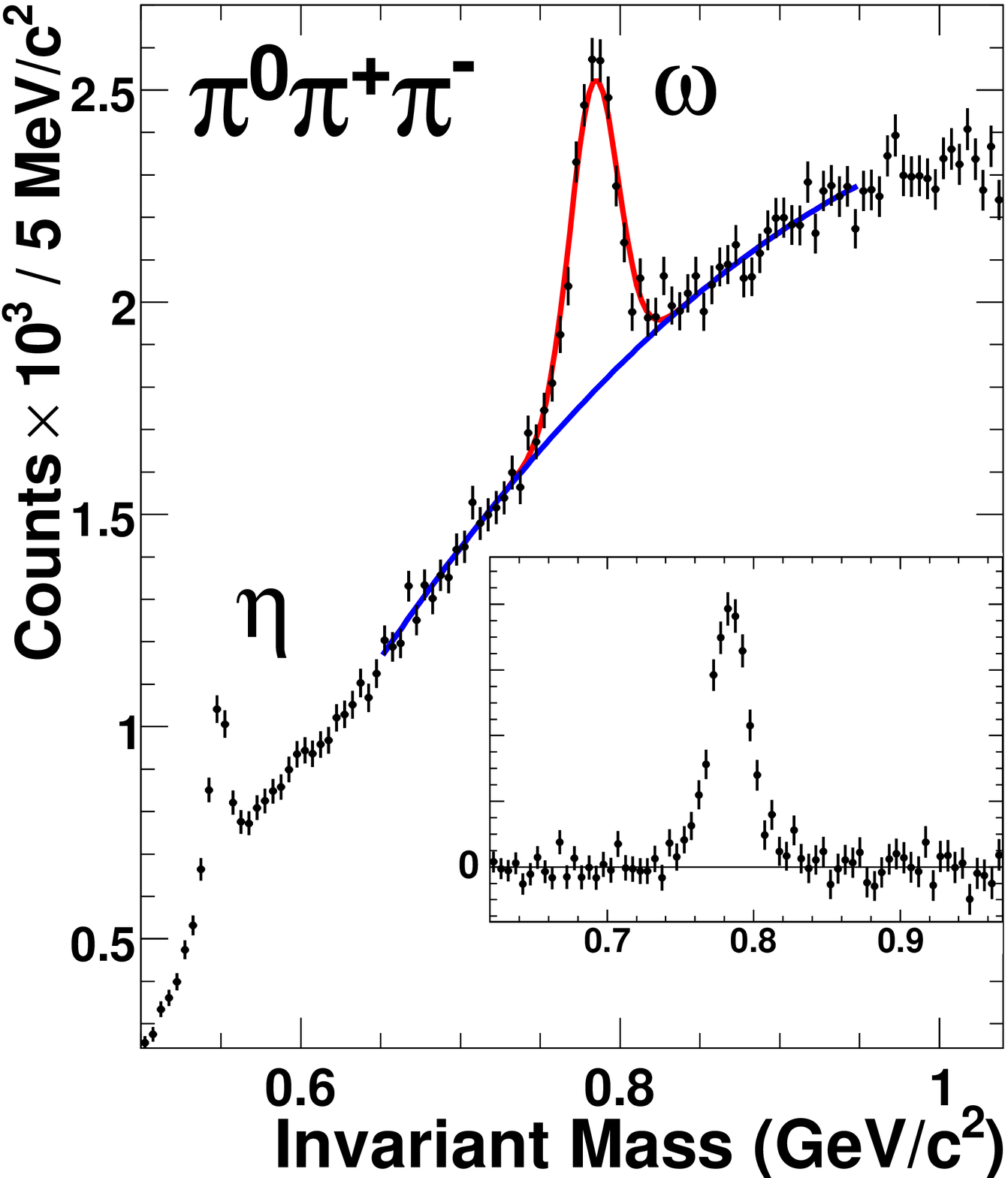} 
\includegraphics[width=0.45\linewidth]{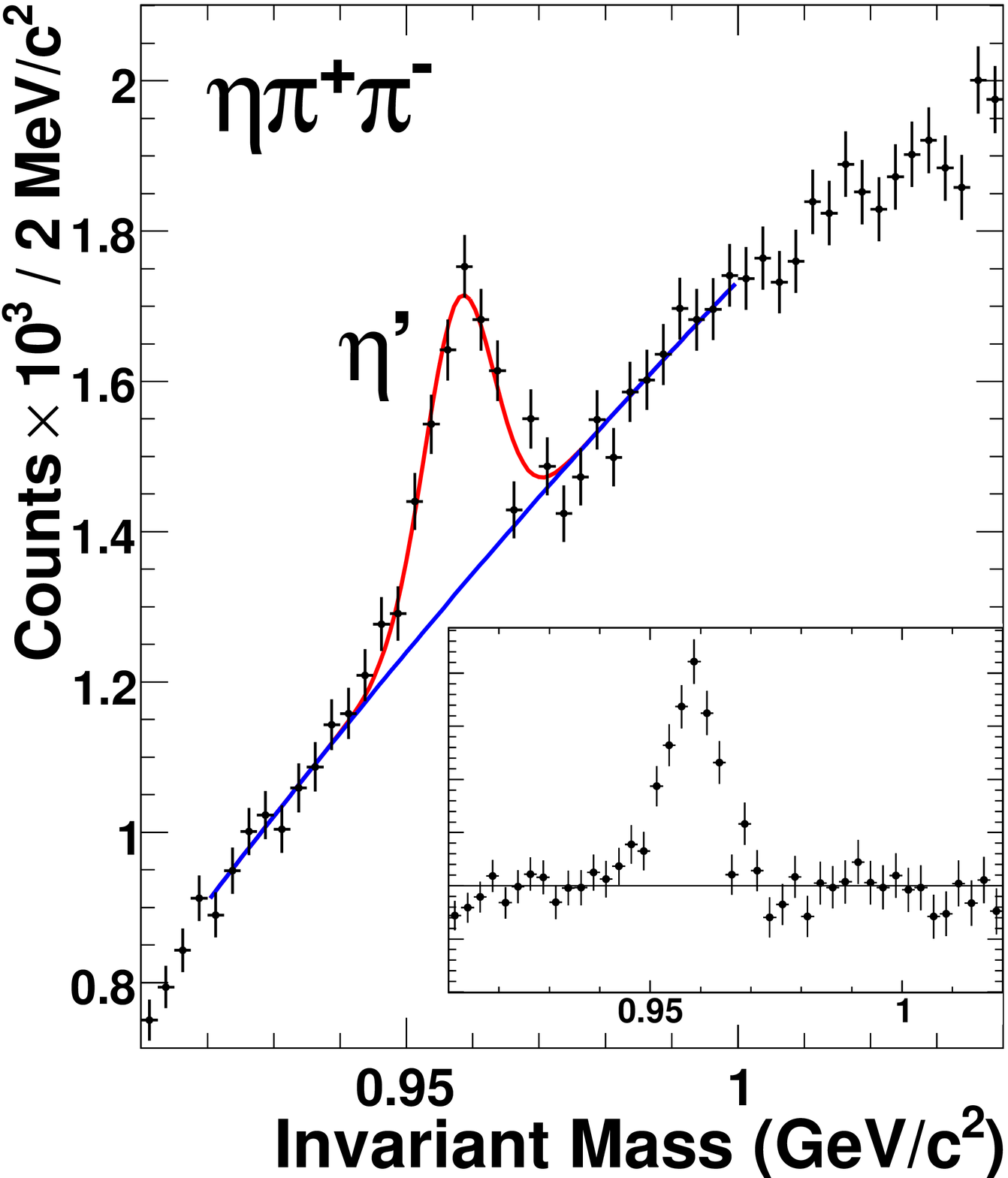}
\caption{Invariant mass distribution for $\pi^{0}\pi^{+}\pi^{-}$ (left) 
and $\eta\pi^{+}\pi^{-}$ (right) triplets in the momentum range 
$4<p_{T}$~(GeV/$c)<6$. The insert shows the invariant mass distribution 
after the background removal explained in sec.~\ref{sec:raw_yield}.
\label{fig:minv_pi0etapippim}}
\end{figure*}

\begin{figure*}[t]
\includegraphics[width=0.45\linewidth]{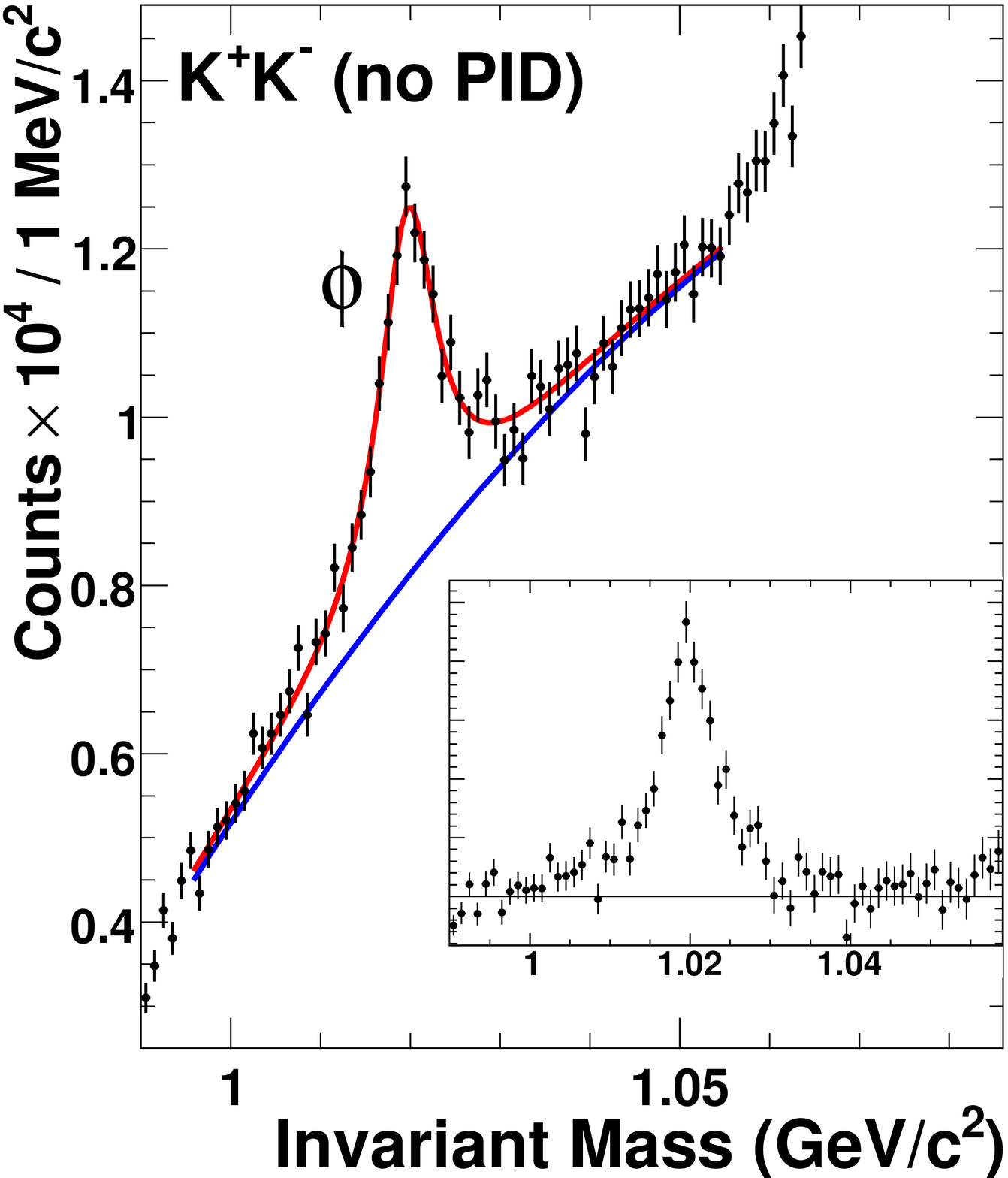} 
\includegraphics[width=0.45\linewidth]{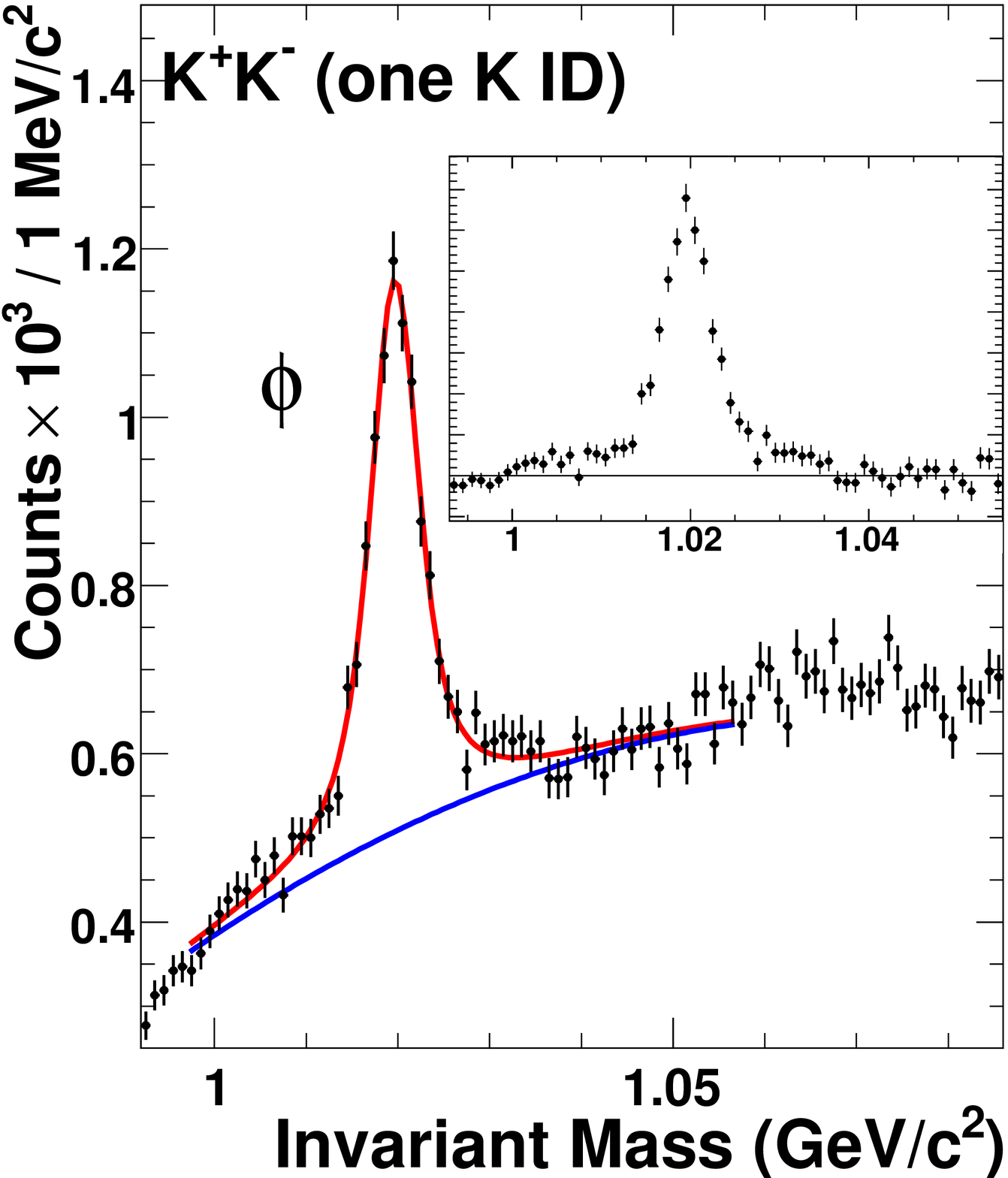}
\caption{Invariant mass distribution for $K^{+}K^{-}$ accumulated without 
particle identification (left) and with one $K$-meson identification 
(right) in the momentum range $4<p_{T}$~(GeV/$c)<6$. The insert shows the 
invariant mass distribution after the background removal explained in 
sec.~\ref{sec:raw_yield}.
\label{fig:minv_KK}}
\end{figure*}

Invariant mass distributions for $\pi^{0}\gamma$ and $\pi^{0}\pi^{0}$
decays are shown in Fig.~\ref{fig:minv_pi0g_pi0pi0}.  The width of the
$\omega$ meson peak is $\sim$~30~MeV and has a weak \pt dependence.
The width of the $K_{S}^{0}$ peak is $\sim$~15~MeV.  The 
signal-to-background ratio (S:B) increases from 1:30 (1:4) to 1:5 (1:2) 
for $\omega$ ($K_{S}^{0}$) mesons as the transverse momentum increases 
from 2 to 12~GeV/$c$.

The main difference in the analysis of the $\omega$ and $K_{S}^0$ decays 
was due to the large lifetime of the $K_{S}^{0}$-meson.  Neutral pions 
coming from the decays of high-\pt $K_{S}^{0}$ originate from a displaced 
vertex and their reconstructed mass and width need to be parameterized in a 
different way compared to pions coming from the primary event vertex. This 
effect was studied using the PHENIX Monte Carlo. The correction was based 
on the mass and width of $\pi^{0}$'s coming from kaon decays with a 
realistic \pt distribution, and on $\pi^{0}$'s produced at the collision 
vertex with the inclusive \pt distribution.
 
\subsubsection{$\omega, \eta \rightarrow \pi^{0} \pi^{+}\pi^{-}$, $\eta' \rightarrow \eta \pi^{+}\pi^{-}$}

For the reconstruction of $\omega, \eta \rightarrow \pi^{0} 
\pi^{+}\pi^{-}$ and $\eta' \rightarrow \eta \pi^{+}\pi^{-}$ decay modes we 
combined $\pi^{0} (\eta)$ candidates with all pairs of oppositely charged 
tracks in the same event~\cite{ppg055,ppg064}. Charged tracks accepted for 
this analysis were required to have momenta in the range 
$0.2<p_{T}$~(GeV/$c)<8$, and were assigned the charged pion mass. Tracks 
with momentum below 0.2~GeV/$c$ do not go through the entire detector due 
to their large bending angle in the axial magnetic field of the central 
magnet.  Tracks that appear to have momenta above 8~GeV/$c$ are, for the 
most part, low momentum secondaries coming from the decay of long lived 
primaries.  Because they do not originate from the collision vertex, their 
momenta are not calculated correctly.  Invariant mass distributions for 
$\pi^{0}(\eta)\pi^{+}\pi^{-}$ triples are shown in 
Fig.~\ref{fig:minv_pi0etapippim}. The two peaks in the distribution shown 
in the left panel of the figure correspond to decays of $\eta$ and 
$\omega$ mesons.  The width of $\sim$~8~MeV/$c^{2}$ for the reconstructed 
$\eta$ meson peak is similar to that of the $\eta'$ meson peak shown in 
the right panel of Fig.~\ref{fig:minv_pi0etapippim}. The width of the 
$\omega$-meson peak is $\sim$~17~MeV/$c^{2}$ which is narrower than that 
in the $\omega \rightarrow \pi^{0}\gamma$ decay mode. This is due to the 
smaller difference between the masses of the primary particle and their 
decay products and to the better momentum resolution of the tracking 
system as compared to the EMCal in this momentum range. The 
signal-to-background ratio in the range of measurements changes from 1:10 
(1:5) to 1:3 (1:2) for $\omega$ ($\eta'$) mesons. More details on the 
analysis of $\eta$ and $\omega$-mesons can be found 
in~\cite{ppg055,ppg064}.

\subsubsection{$\phi \rightarrow K^{+}K^{-}$}

Reconstruction of the $\phi \rightarrow K^{+}K^{-}$ decay was done by 
combining pairs of oppositely charged tracks.  The tracks were required to 
have a momentum in the range $0.3<p_{T}$~(GeV/$c)<8$. Each track was 
assigned the charged kaon mass. Invariant mass distributions were 
accumulated in two different configurations: i) combining all tracks 
reconstructed in the PHENIX tracking system; ii) combining all tracks of 
one sign with tracks of the opposite sign identified as a kaon in the TOF 
subsystem.  Examples of the invariant mass distributions for the two cases 
are shown in the left and right panels of Fig.~\ref{fig:minv_KK}, 
respectively.

The use of particle identification improved the signal-to-background ratio 
by a factor of more than two at the expense of a more limited acceptance, 
resulting in a factor of five loss in statistics. At low and intermediate 
\pt, where the combinatorial background is high but the data sample has 
large statistics this method is preferable. The method without particle 
identification was more effective at intermediate and high \pt because of 
the significant gain in the acceptance.  The highest \pt reachable with 
this method is limited by the available statistics in the minimum bias data 
sample. The two methods described here use different detector subsystems 
and produce different shapes of combinatorial background and 
signal-to-background ratios. Use of the two methods allowed to extend the 
\pt coverage of the measurement and provided a consistency check between 
the results obtained in the overlap region between 1.5~GeV/$c$ and 
4.5~GeV/$c$. The signal-to-background ratio changes from 1:10 to 2:1 
depending on the analysis method and the \pt bin. More details on this 
measurement can be found in~\cite{ppg096}.


\begin{figure}[htb]
\includegraphics[width=1.0\linewidth]{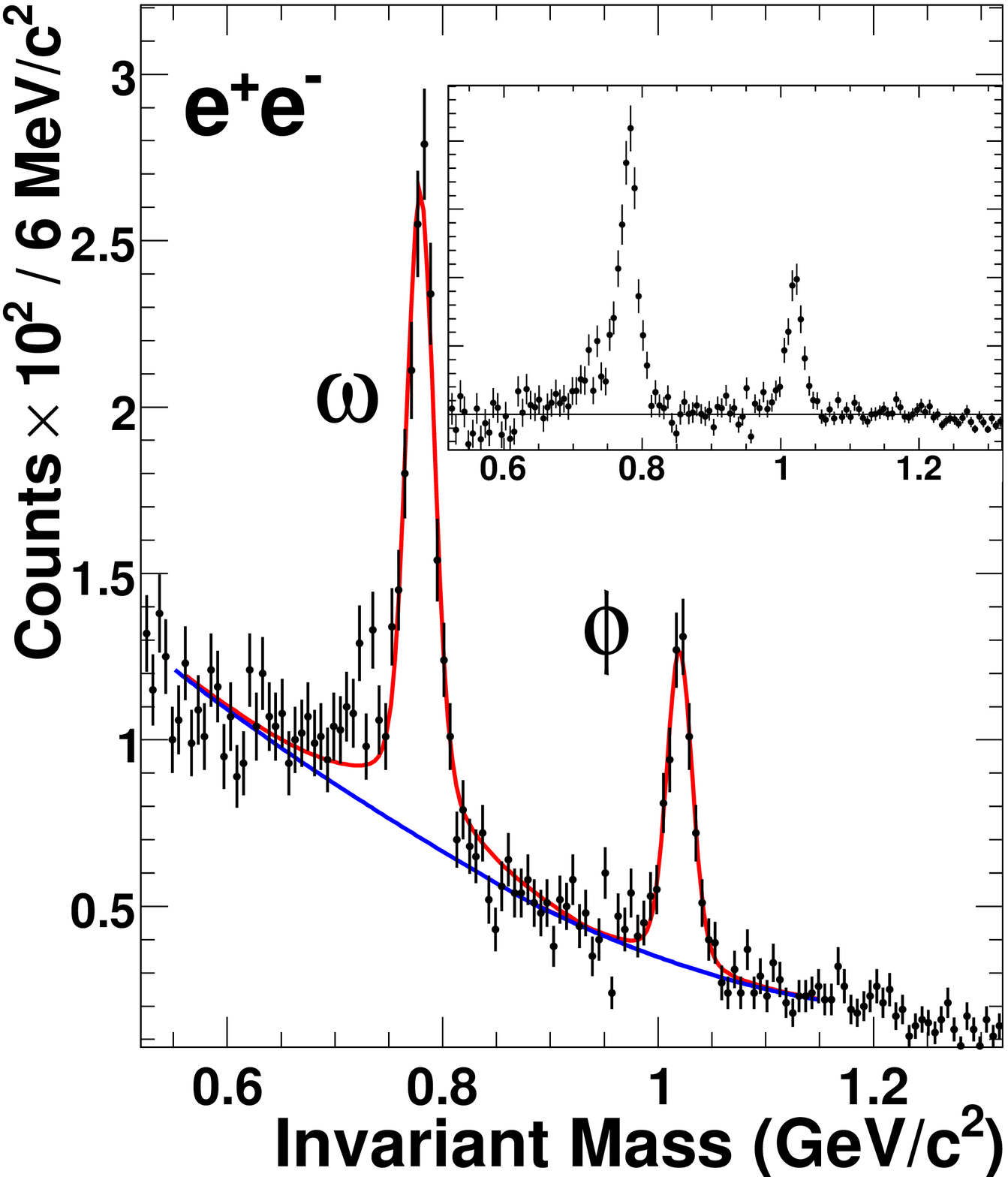}
\caption{Invariant mass distribution for $e^{+}e{-}$ pairs in the momentum 
range $0.5<p_{T}$~(GeV/$c)<0.75$. The insert shows the invariant mass 
distribution after the background removal explained in 
sec.~\ref{sec:raw_yield}.
\label{fig:om_phi_ee}}
\end{figure}

\subsubsection{$\omega, \phi \rightarrow e^{+}e^{-}$}

Electrons are reliably identified by the PHENIX detector in the momentum 
range $0.2<p_{T}$~(GeV/$c)<4$. Electron identification is accomplished 
using the information from the RICH and EMCal subsystems by requiring at 
least two RICH phototubes to fire within the ring shaped area associated 
with a charged track. In addition, the ratio of the associated cluster 
energy measured in the EMCal to the momentum measured in the tracking 
system must satisfy $\left|E/p-1\right|<0.5$. The invariant mass 
distribution obtained by combining identified $e^+$ and $e^-$ pairs is 
shown in Fig.~\ref{fig:om_phi_ee} for pairs in the range 
$0.5<p_{T}$~(GeV/$c)<0.75$. The two peaks correspond to $\omega + \rho$ 
and $\phi$ mesons. The widths of the $\omega$ $(\phi)$ meson peaks vary 
from 6.1(6.0)~MeV/$c^{2}$ to 9.0(11)~MeV/$c^{2}$ from the lowest $p_{T}$ 
to the highest $p_{T}$ of the electron pairs. The signal-to-background 
ratio in the region of the $\omega$ $(\phi)$ meson peaks changes from 
1:2(2:1) to 3:1(6:1).

\subsection{Raw yield extraction \label{sec:raw_yield}}

To extract the raw yields the invariant mass
distributions near each peak were parameterized as the
sum of signal and background contributions.

For the signal, we used a Breit-Wigner function convolved with a Gaussian 
function (\bwg). The Breit-Wigner describes the natural shape of the 
measured resonance and the Gaussian takes into account the detector 
resolution.  Depending on the decay channel being analyzed, one or the 
other contribution may dominate, e.g. the Gaussian part is more important 
in decays like $\omega \rightarrow \pi^{0}\gamma$, or $K_{s}^{0} 
\rightarrow \pi^{0}\pi^{0}$, and the Breit-Wigner part in decays like $\phi 
\rightarrow K^{+}K^{-}$ or $\omega, \phi \rightarrow e^{+}e^{-}$.

In most cases the parameters of the \bwg function when fitted to the data 
were consistent with the values expected from simulation.  In the highest 
\pt bins, where the available statistics becomes a limiting factor, we 
constrained the Gaussian width based on simulations.

The $\phi \rightarrow K^{+}K^{-}$ decay mode was treated somewhat 
differently. Kaons decaying in flight before passing completely through the 
PHENIX tracking system modify the shape of the invariant mass distribution 
compared to those passing through the detector without decays.  This 
results in non-Gaussian tails of the detector response function, and thus 
the Breit-Wigner and Gaussian width parameters in the \bwg convolution mix 
together.  To account for this effect a Monte Carlo sample was produced 
with the natural width of the $\phi$ set to zero and the kaon lifetime set 
to infinity.  Using these samples allowed to disentangle the effects 
related to the kaon decays in flight.

In the analysis it was verified that the peak positions and widths obtained 
from the fits to the data were in agreement with the simulated data to 
within 10\%. In the measurement of the $\omega,\phi \rightarrow e^{+}e^{-}$ 
decays other terms were added to the \bwg shape to account for $\rho$ 
decays and for internal conversions taken 
from~\cite{rad_tail_1,rad_tail_2}. The contribution of $\rho$ underneath 
the $\omega$ peak was estimated using Breit-Wigner parameterization, with 
the assumption that the production ratio of $\rho$ and $\omega$ is 1 and in 
the fit their ratio was determined by their $e^+e^-$ branching ratios in 
vacuum equal to 1.53.

To properly estimate the background under the peak it is necessary to 
assume that the shape of the background does not change rapidly. With this 
assumption one can expand the background shape in a Taylor series around 
the peak position and take the most significant terms of the expansion. A 
natural choice is to use a second order polynomial. The regions outside the 
resonance peak, where the background dominates, define the parameters of 
the fit. For a second order polynomial fit the background varies smoothly 
under the peak. This may not be the case for higher order polynomial fits 
to the background.

The combinatorial background in the data has two main contributions. The 
first comes from the random association of uncorrelated tracks. Its shape 
is defined by the detector acceptance and the \pt distribution of particles 
in the event. This part of the background remains smooth in the mass 
interval comparable to the width of the peaks shown in 
Figs.~\ref{fig:minv_pi0g_pi0pi0}-\ref{fig:om_phi_ee}.  The correlated part 
of the combinatorial background comes from partially or incorrectly 
reconstructed decays of true particles and jets, and may have a faster 
changing shape. In several analyses the most significant contributions to 
the correlated background were studied to verify that they do not affect 
the raw yield extraction procedure. For example, the decay $\eta 
\rightarrow \gamma\gamma$ produces an $\eta\gamma$ peak at around 
0.6~GeV/$c^{2}$ in the invariant mass distribution of $\pi^{0}\gamma$. 
Also, the decay of $K_{s}^{0} \rightarrow \pi^{+}\pi^{-}$ produces a peak 
at $\sim$1.07~GeV/$c^{2}$ in the $K^{+}K^{-}$ invariant mass distribution 
when two pions are erroneously assigned the kaon mass.  In some cases these 
processes limit the mass range available for the background determination. 
The mass range used for determination of the background did not include 
regions where one could expect appearance of such peaks.

The raw yields were measured in the following way. First, the invariant 
mass distributions in different \pt bins were fitted with the \bwg plus 
background in the mass range of $\pm$5 combined widths of the \bwg around 
the nominal mass of the meson. The exact range varied slightly depending on 
particle species and the \pt bin. The background contribution, estimated by 
the second order part of the fit function, was subtracted from the measured 
invariant mass distribution and the resulting histogram was used to count 
the raw yield. Bins lying within $\pm$2.5 combined widths of the \bwg 
function around the mass peak contributed to the yield. The same procedure 
was used to calculate the raw yield in the Monte-Carlo used for the 
acceptance evaluation.

The systematic uncertainty of the raw yield extraction was usually the main 
contributor to the total systematic uncertainty.  We evaluated this 
uncertainty by modifying the analysis procedure. The main goal of that was 
to change the shape of the background around the resonance peak in a manner 
similar to that shown in Fig.~\ref{fig:minv_KK}. To achieve this goal, 
analyses of the same decay modes were performed in different ways. For 
example, by requiring PC3 or EMCal hit matching for charged tracks, varying 
the minimum energy of $\gamma$ clusters, or modifying the selection 
criteria for $\pi^{0}$($\eta$) candidates.  Independent of that we also 
varied the parameters of the fit functions, such as the fit range and the 
order of the polynomial.  Typically, six to ten raw yield values were 
accumulated for each \pt bin.  After fully correcting each of them for the 
corresponding reconstruction efficiency the variance of the results was 
taken as the systematic uncertainty.

\subsection{Invariant mass resolution}

\begin{figure}[htb]
\includegraphics[width=1.0\linewidth]{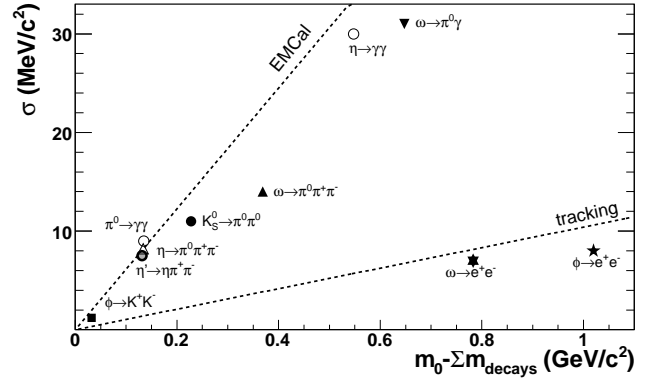}
\caption{Invariant mass resolution of the PHENIX detector for different 
decay modes measured in the momentum range $4<p_{T}$~(GeV/$c)<6$, except 
for the $e^{+}e^{-}$ mode that is measured in the range 
$0.5<p_{T}$~(GeV/$c)<0.75$. The lines indicate the expected detector mass 
resolution.
\label{fig:mass_res}}
\end{figure}

The invariant mass resolution of the detector plays an important role in 
the analyses described in this section. It depends on several factors. Use 
of the detector tracking system or EMCal makes a large difference. The 
momentum range of the analyzed particles is less important. The difference 
between the mass of the particle and its decay products contributes 
directly to the invariant mass resolution. To demonstrate this we consider 
the limiting case of a particle decaying into two massless products. In 
this case, one can approximate the invariant mass resolution with the 
simple relation $\delta m/m = (1/\sqrt{2})\delta p_{T}/p_{T}$. The single 
particle momentum resolution was discussed in section~\ref{sec:detector}. 
Figure~\ref{fig:mass_res} compares this approximation with the widths of 
the peaks shown in Figs.~\ref{fig:pi_eta_gg}~-~\ref{fig:om_phi_ee}. The 
measured widths are plotted as a function of the mass difference between 
the particle and its decay products. The two lines in the plot are 
calculated for two body decays reconstructed either with the tracking 
system only, or with the EMCal only at pair \pt of 4~GeV/$c$.

As can be seen, the simple approximation describes the measured mass widths 
for the two body decays reasonably well. The widths of the $e^{+}e^{-}$ 
decay modes are somewhat narrower due to use of a lower momentum range. The 
results for the $J/\psi$ and $\psi'$ which are not shown in the plot are 
also consistent with the trend of the ``tracking'' line. The 
$\phi\rightarrow K^{+}K^{-}$ represents the case where the assumption of 
massless products is least valid, nevertheless, the agreement is still 
reasonable.

The widths of the invariant mass peaks reconstructed with both the EMCal 
and the tracking systems are dominated by the EMCal resolution. However, 
due to the energy correction applied to the $\gamma$ clusters forming 
$\pi^{0}$ or $\eta$ candidates the widths of the peaks reconstructed with 
3 and 4 particles are below the ``EMCal'' line.

\subsection{Detector acceptance and efficiency}

\subsubsection{Geometrical acceptance and the analysis cuts}

The determination of the detector acceptance was done using a single 
particle Monte Carlo simulation. Particles were uniformly generated within 
$\left|y\right|<0.5$ in rapidity and in full azimuthal angle. The range of 
the transverse momentum distributions were chosen to produce sufficient 
statistics in all \pt bins for which the signal could be extracted from 
the data. For the acceptance calculation the generated spectra were 
weighted to match the measured particle spectra. This procedure was done 
iteratively. Kinematics of the three-body decays of the $\eta$, $\omega$, 
and $\eta'$ mesons assumed the experimentally measured phase space density 
distributions~\cite{omega_dalitz_1,omega_dalitz_2,omega_dalitz_kloe,omega_dalitz_3,etap_dalitz_1,etap_dalitz_2}.

A GEANT based simulation of the PHENIX detector was tuned to reproduce the 
response of all detector subsystems and inactive areas. It was verified 
that the simulated positions and widths of the $\pi^{0}$, $K_{s}^{0}$, 
$\eta$, $\omega$, $\eta'$, and $\phi$ peaks were consistent with the 
values measured in real data at all \pt's. The same analysis code was used 
for the reconstruction and analysis of the simulated and real data.

\begin{figure*}[htb]
\includegraphics[width=0.4\linewidth]{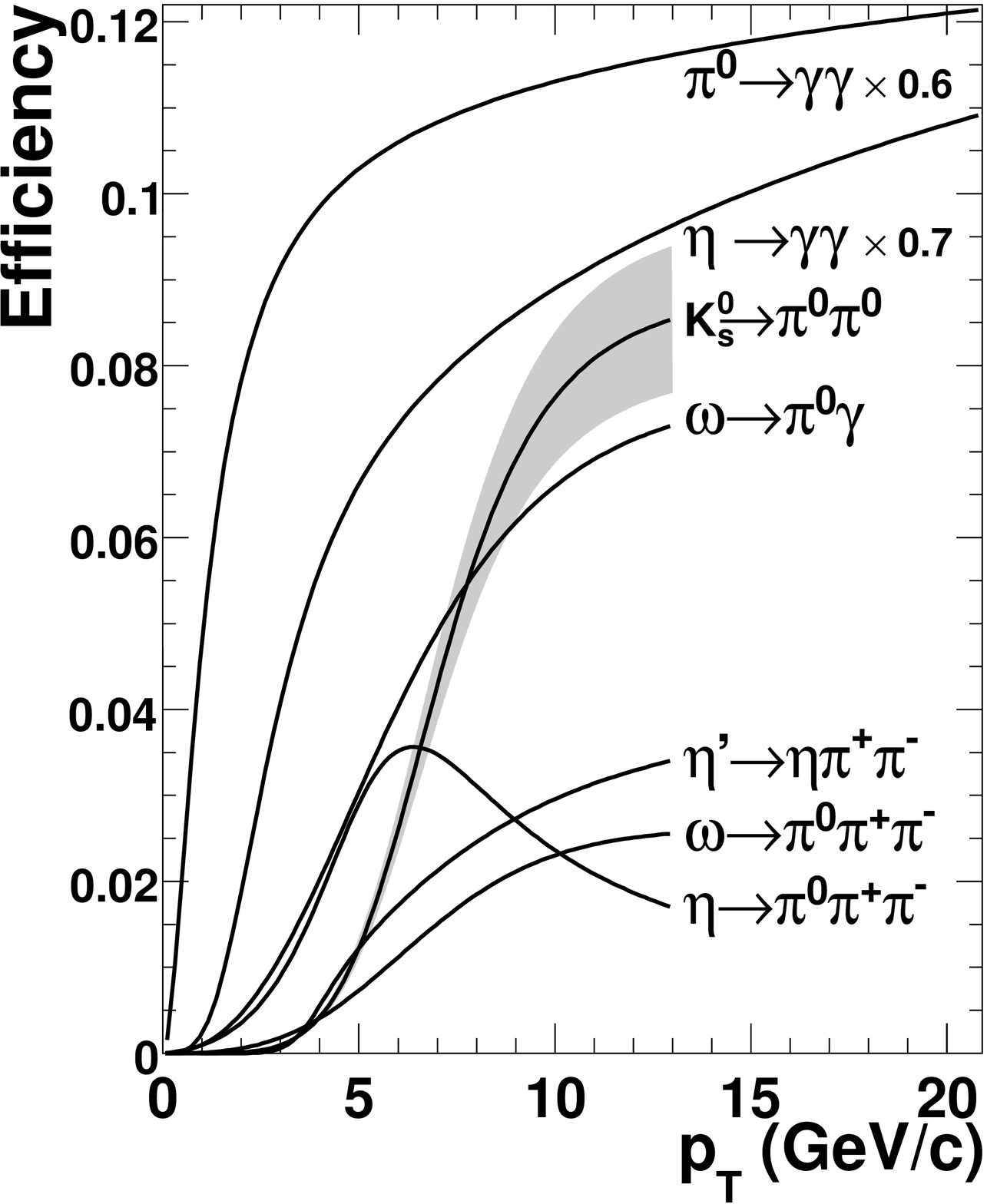}
\includegraphics[width=0.4\linewidth]{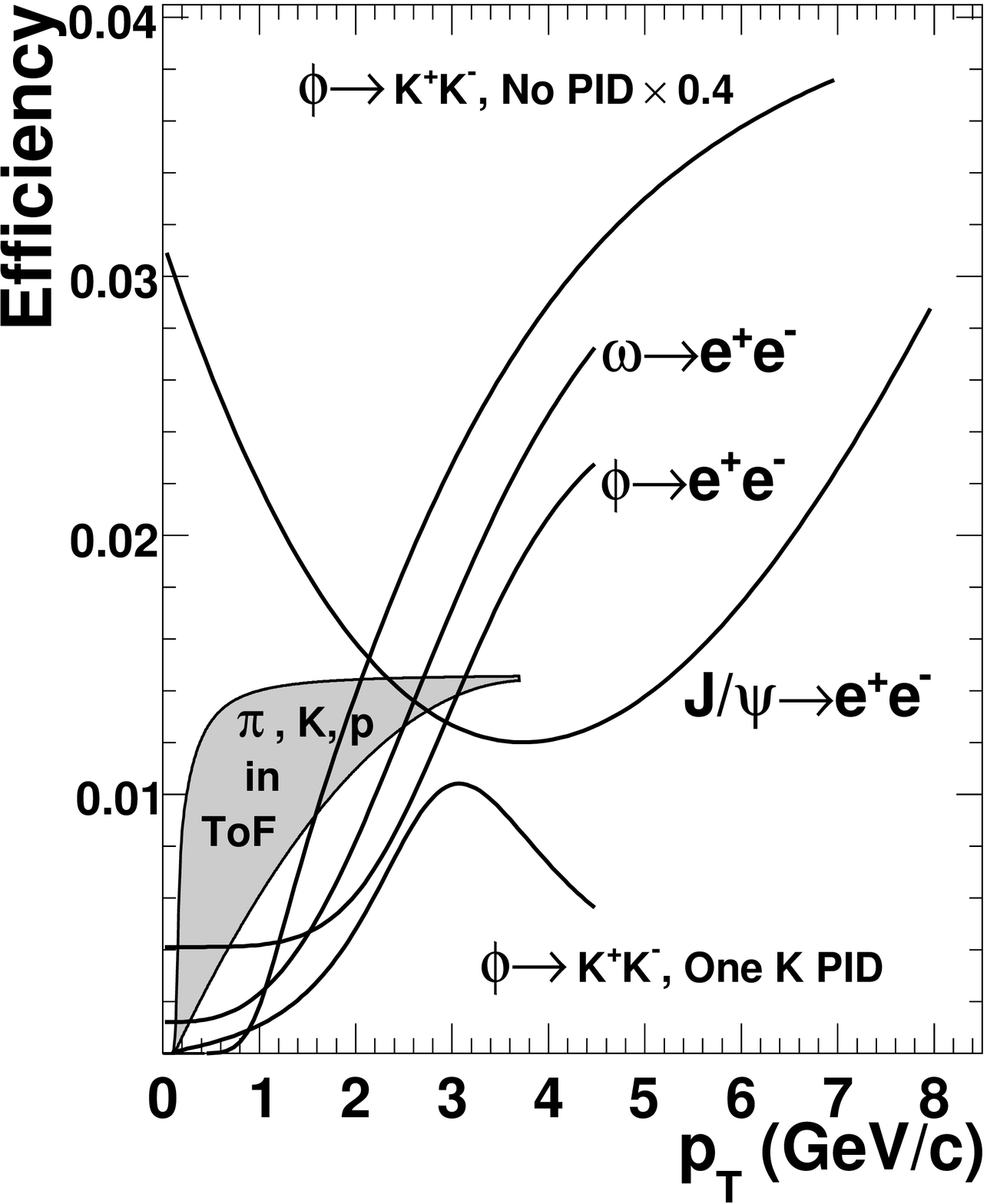}
\caption{Detector acceptance as a function of transverse momentum for 
different particles measured with the PHENIX experiment. The band in the 
left panel shows the largest relative systematic uncertainty among all 
curves.
\label{fig:acc_all}}
\end{figure*}

The detector acceptance, calculated as the ratio of the number of fully 
reconstructed particles to the number of generated particles, is shown in 
Fig.~\ref{fig:acc_all}. All curves take into account the detector 
geometry, particle decay kinematics, performance of the detector 
subsystems including particle identification, and the analysis cuts. The 
efficiencies strongly depend on the particle momentum and rapidly decrease 
at low \pt for all species studied in this analysis, establishing a low 
\pt edge for the measurements.

\begin{figure*}[htb]
\includegraphics[width=0.4\linewidth]{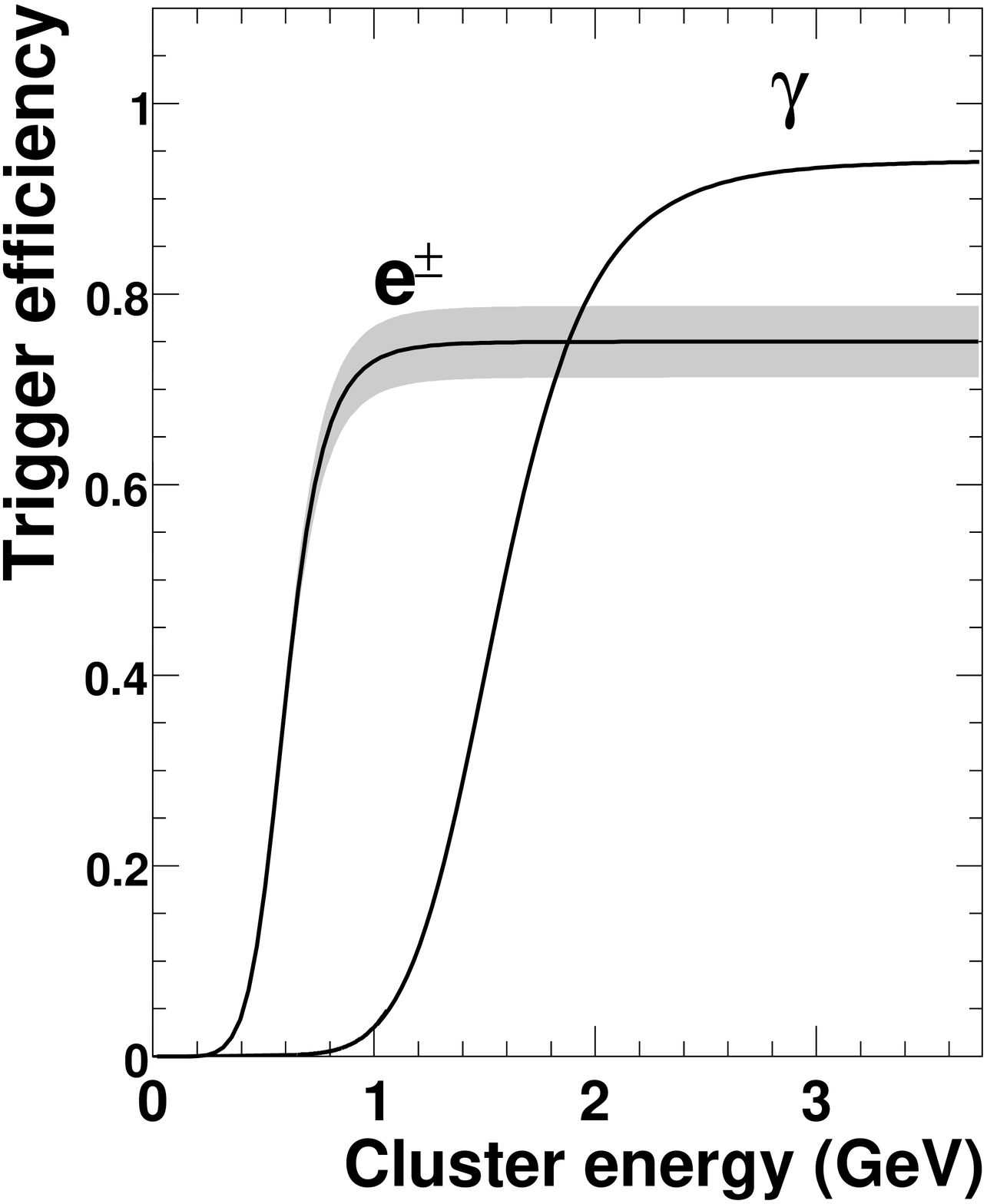}
\includegraphics[width=0.4\linewidth]{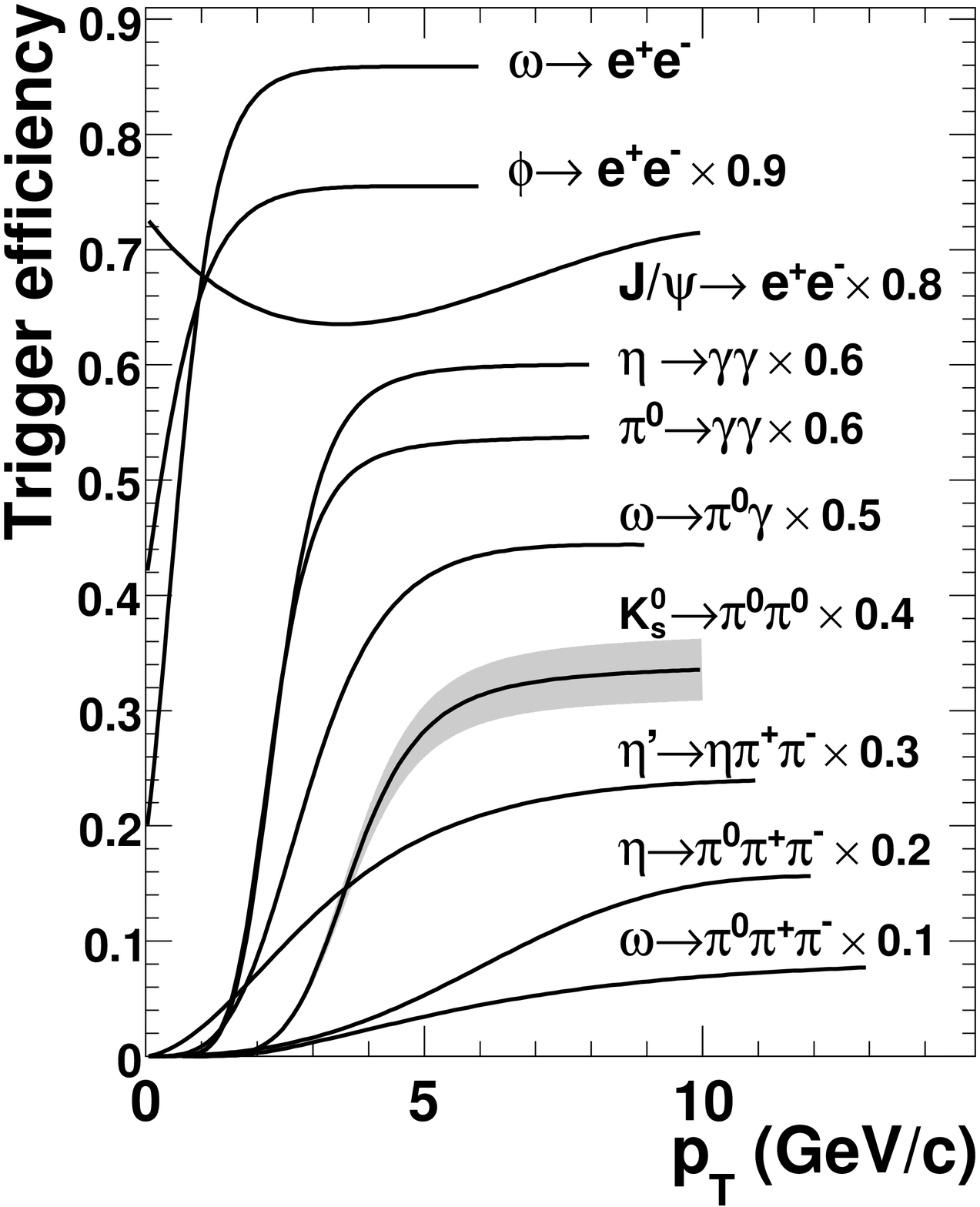}
\caption{Efficiencies of the ERT 4x4 and ERT 2x2 triggers for single
  $\gamma$ clusters and electrons as a function of energy (left).
  Trigger efficiencies for different meson decays as determined
  from Monte Carlo simulation (right). The band shows the largest relative 
systematic uncertainty among all curves. \label{fig:treff_all}}
\end{figure*}

\subsubsection{ERT trigger efficiency}

The analysis of several decay modes was based on data samples accumulated 
with the ERT trigger described in section~\ref{sec:detector}. The ERT 
trigger efficiency was extracted using the minimum bias event sample. Each 
EMCal cluster which set the ERT trigger bit to indicate a $\gamma$-cluster 
or electron was identified.  The track or cluster had to also satisfy the 
analysis cuts of a particular decay mode, and match the region where the 
trigger bit was generated.  The trigger efficiency was calculated as the 
energy spectra of such clusters divided by the energy spectra of all 
accepted clusters or electrons.  Trigger efficiencies of photons and 
electrons measured for one of the PbSc sectors as a function of cluster 
energy are shown in the left panel of Fig.~\ref{fig:treff_all}.

The trigger efficiencies grow steeply with energy, reaching 50\% at values 
approximately corresponding to the on-line trigger threshold setting of 
0.6~GeV for electrons and 1.4~GeV for photons.  The curves saturate at 
approximately twice the threshold energy. The level of saturation is below 
100\% because of inactive areas of the ERT and the RICH efficiency.

For the analyzed decay modes the trigger efficiency evaluation was done 
using the same Monte-Carlo sample as was used for the acceptance 
calculation. First we required the particle to be reconstructed in PHENIX 
without ERT trigger requirement. Then, for all EMCal clusters associated to 
photons or electrons in the final state of the decay, we generated a random 
number between 0 and 1 and compared it to the magnitude of the curve shown 
in the left panel of Fig.~\ref{fig:treff_all} at the energy of the cluster. 
The particle was considered to fire the ERT trigger if at least one of the 
randomly generated numbers was lower than the corresponding value of the 
curve. The probability to fire the ERT trigger for all analyzed mesons is 
shown in the right panel of the same figure.

\subsubsection{Electron identification efficiency}

The electron identification efficiency is included in the acceptance 
efficiencies shown in Fig.~\ref{fig:acc_all}. It was evaluated 
using a full detector Monte Carlo simulation which was tuned to 
adequately reproduce the RICH and the EMCal detector responses. To ensure 
that the electron identification efficiency was properly done in the 
simulation it was confirmed to agree with the efficiency measured with real 
data.

\begin{figure}[htb]
\includegraphics[width=1.0\linewidth]{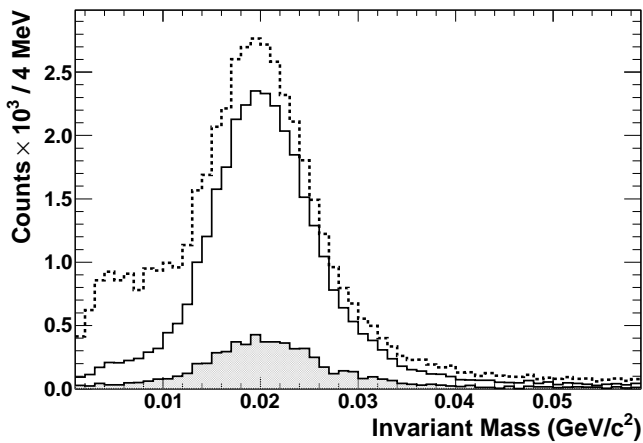}
\caption{Invariant mass distribution for $e^{+}e^{-}$ pairs where one 
track is identified as an electron and the second track is any track 
(dashed line), the same for pairs which open up in the plane perpendicular 
to the detector magnetic field (solid line), and among those pairs, the 
ones in which the second track fails the electron identification cut 
(filled histogram).
\label{fig:eID}}
\end{figure}

For this comparison the data samples accumulated during special PHENIX runs 
were used. In those runs a 1.7\% radiation lengths brass converter was 
installed around the RHIC beam pipe in the PHENIX interaction region. In 
this sample we selected electrons of both signs using very strict electron 
identification requirements. Those electrons were paired with all other 
tracks in the event. The invariant mass distribution of such pairs is shown 
by the upper histogram in Fig.~\ref{fig:eID}.

One can see the characteristic shape of the partially reconstructed 
$\pi^{0}$ Dalitz decays and a peak at around 22~MeV/$c^{2}$ corresponding 
to $\gamma$-conversions close to the beam pipe. Since the conversion 
electrons originate at the displaced converter vertex, and therefore skip 
the first 3.8~cm of the magnetic field, the reconstructed invariant mass 
peak is shifted from zero. Among these pairs a further selection was made 
to choose those which open up in the plane perpendicular to the detector 
magnetic field. This requirement effectively suppresses the combinatorial 
background and pairs coming from the $\pi^{0}$ Dalitz decays, but keeps 
$\gamma \rightarrow e^{+}e^{-}$ pairs having small opening angle. The 
middle histogram in Fig.~\ref{fig:eID} shows that the conversion peak 
significantly dominates the residual Dalitz contribution and the 
combinatorial background. Finally we applied the electron identification 
requirements to the second track. The invariant mass distribution of the 
pairs where the second track fails to be identified as an electron is shown 
by the filled histogram. The ratio of the lowest to the middle histogram 
under the peak is the electron identification loss. It reaches 20\% below 
0.5~GeV/$c$ and saturates at $\sim$~10\%.

\subsection{Calculation of invariant cross sections}
The invariant cross section for a particle in each \pt bin was calculated as:
\begin{equation}
\frac{1}{2\pi \pt} \frac{d^2\sigma}{d\pt dy} = \frac{1}{2\pi \pt}
\frac{1}{\mathcal{L} \, {\rm BR}} \,\frac{1}{\varepsilon(\pt) \, \varepsilon_{\rm BBC}}
\frac{N(\Delta \pt)}{\Delta \pt \, \Delta y}
\label{eq:invyld}
\end{equation}
where $N(\Delta \pt)$ is the number of reconstructed particles in a given 
\pt bin, $\mathcal{L}$ is the integrated luminosity sampled by the minimum 
bias trigger, $\varepsilon(\pt)$ is the acceptance and reconstruction 
efficiency, $BR$ is the branching ratio, and $\varepsilon_{\rm BBC}$ is 
the minimum bias trigger efficiency for events containing mesons, 
estimated to be $0.79\pm0.02$. The cross section sampled by the BBC 
trigger, $\sigma_{\rm tot}^{pp} = 23.0 \pm 2.2$~mb, was used to determine 
the integrated luminosity. For the analyses with the minimum bias data 
sample $\varepsilon(p_{T})$ corrects for the acceptance and reconstruction 
efficiency while for analyses with the ERT data sample it includes the ERT 
trigger efficiencies as well. A bin shift correction was applied to take 
into account the finite width of the \pt bins used in the analyses. The 
correction is made by shifting the data points along the vertical axis 
according to the procedure described in~\cite{binshift}.

Finally, in the $\omega\rightarrow\pi^{0}\pi^{+}\pi^{-}$ and $K_{S}^{0} 
\rightarrow \pi^{0}\pi^{0}$ analyses, the cross sections measured with the 
ERT and with the minimum bias triggers were averaged in the overlapping \pt 
region, taking into account the statistical and systematic uncertainties.

\subsection{Systematic uncertainties \label{sec:errors}}

In addition to the systematic uncertainties described in the corresponding 
analysis sections, uncertainties of the ERT trigger efficiency and 
acceptance corrections were estimated by varying the analysis cuts, and by 
varying the energy and momentum scales of the EMCal and DC by 1\%. The 
resulting systematic uncertainties for the different decay modes of 
$K_{s}^{0}$, $\eta$, $\omega$, $\eta'$, and $\phi$ mesons are summarized in 
Table~\ref{tab:tabsyst}. The uncertainties are categorized by types: (A) 
uncorrelated between \pt bins, (B) \pt correlated, all points move in the 
same direction but not by the same factor, (C) an overall normalization 
uncertainty in which all points move by the same factor independent of \pt. 
The type C uncertainty is predominantly due to the uncertainty of the 
minimum bias trigger efficiency in \pp collisions, equal to 
9.7\%~\cite{ppg030,ppg055}. The uncertainty of the raw yield extraction is 
estimated as described in subsection~\ref{sec:raw_yield}. It dominates the 
total uncertainty and is split into Type A and Type B contributions.

\begin{table*}[htb]
\caption{Relative systematic uncertainties (in percent) for different decay modes. Given ranges indicate the variation of the systematic 
uncertainty over the \pt range of the measurement. \label{tab:tabsyst}}
\begin{ruledtabular}
\begin{tabular}{lcccccccc}
Particle &$K_{s}^{0}$      &           \multicolumn{3}{c}{$\omega$}            &  $\eta\prime$      &\multicolumn{2}{c}{$\phi$}  &  Uncertainty \\
Decay    &$\pi^{0}\pi^{0}$&$\pi^{0}\pi^{+}\pi^{-}$&$\pi^{0}\gamma$&$e^{+}e^{-}$&$\eta\pi^{+}\pi^{-}$&$K^{+}K^{-}$&$e^{+}e^{-}$   &  Type \\
\hline
Acceptance	         & 8	&5	&6	&5	&5	&5-7	&5      & B \\
EMCal energy resolution	 & 4-5	&2-5	&2-3	&	&2-4	&	&	& B \\	
EMCal, DC scale          & 4-6	&2-6	&3-17	&2-11	&2-5	&1-5	&2-10	& B \\
$\pi^{0},\eta$ selection & 5-10	&3	&3	&	&3	&	&	& B \\	
ERT trigger efficiency	 & 2-12	&3-10	&2-7	&1-3	&2-4	&	&1-2    & B\\
Peak extraction MC	 & 2	&1	&1	&1	&1	&3	&1	& A,B\\
Raw yield extraction	 & 4-19	&5-17	&5-12	&4-15	&6-25	&8-25	&3-11   & A,B\\
$\gamma$-conversion	 & 6	&3	&5	&	&3	&	&	& C\\
$e$-identification	 & 	& 	& 	&10	&	&	&9	& B\\
Branching ratio	         & 0	&1	&3	&1.7	&3	&1	&1.3    & C\\
MinBias Trigger	         &9.7	&9.7	&9.7	&9.7	&9.7	&9.7	&9.7    & C\\
\hline			 	 	 	 	    	     	 	 	         	 
Total	                 &17-29	&13-24	&15-26	&16-24	&14-29	&14-28	&15-18  &\\
\end{tabular}
\end{ruledtabular}
\end{table*}

\subsection{Neutral meson spectra}

\begin{figure}[htb]
\includegraphics[width=1.0\linewidth]{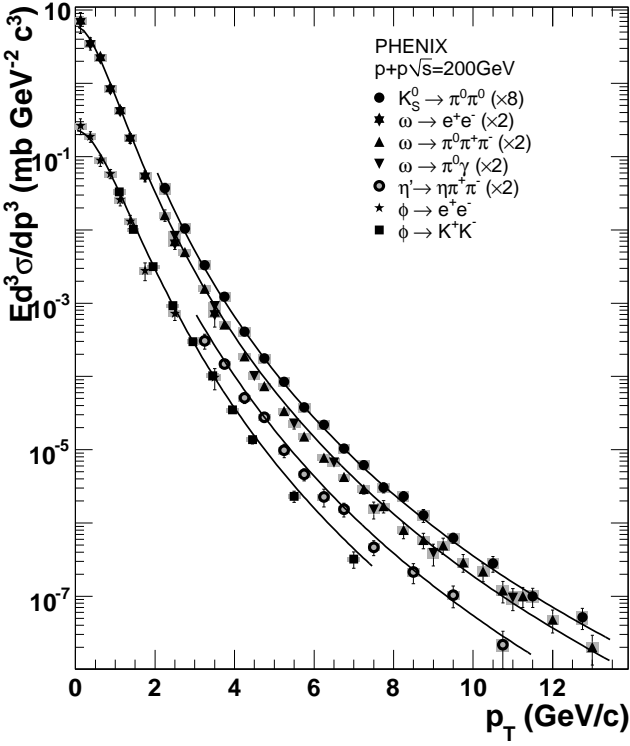}
\caption{Invariant differential cross-section of neutral mesons measured in \pp
collisions at \sqs=200~GeV in various decay modes. The lines are fits to the
spectra as described further in the text.
\label{fig:new_tranc}}
\end{figure}

The invariant differential cross sections calculated using 
Eq.~\ref{eq:invyld} are tabulated in 
Tables~\ref{tab:omega}~and~\ref{tab:mesons} and plotted in 
Fig.~\ref{fig:new_tranc}. Different symbols are used to show results for 
different decay modes. One can see a very good agreement between the 
particle spectra measured in the different decay modes. Results for low \pt 
bins for particles reconstructed through decays in the $e^{+}e^{-}$ mode 
allow an accurate measurement of the integrated particle yield.  The 
integrated yield at midrapidity for the $\omega$ is measured to be 
$d\sigma^{\omega}/dy=4.20\pm0.33^{\rm stat}\pm0.52^{\rm syst}$~mb and for 
the $\phi$ is measured to be $d\sigma^{\phi}/dy=0.432\pm0.031^{\rm 
stat}\pm0.051^{\rm syst}$~mb. The mean transverse momentum for these 
particles is $\langle p_{T}^{\omega}\rangle=0.664\pm0.037^{\rm 
stat}\pm0.012^{\rm syst}$~GeV/$c$ and $\langle 
p_{T}^{\phi}\rangle=0.752\pm0.032^{\rm stat}\pm0.014^{\rm syst}$~GeV/$c$.

\section{Analysis of particle spectra \label{sec:ana_scaling}}

In this section we analyze the measured invariant transverse momentum 
spectra for a variety of hadrons in \pp collisions at \sqs = 200~GeV and 
search for common features. All measurements are quoted as the invariant 
differential cross sections at midrapidity averaged over the rapidity 
interval $|y| \le 0.35$.
\begin{eqnarray}
E\frac{d^{3}\sigma}{dp^{3}} = \sigma_{pp}^{{\rm inel}}\times\frac{1}{2\pi
  p_T}\frac{1}{N_{events}}\frac{d^{2}N}{dy\,dp_T}
\label{eq:cs}
\end{eqnarray}
where $\sigma_{inel}^{pp}$ = 42~mb.

\subsection{Data samples \label{sec:data_samples}}

The procedures used for the reconstruction of the particle transverse 
momentum spectra are described above in section~\ref{sec:ana_part} and in 
other PHENIX publications listed in Table~\ref{tab:data_samples}. 
Figure~\ref{fig:new_tranc} shows (left) the results and (right) the results 
compared with previous PHENIX results. All meson spectra used in this paper 
are not corrected for feed down.

\begin{table}[htb]
\caption{Data samples used in the analysis of particle spectra.
The~\protect\ref{tab:omega} and \protect\ref{tab:mesons} in the 
``Ref." column refer to Appendix Tables~\protect\ref{tab:omega} 
and \protect\ref{tab:mesons}.
\label{tab:data_samples}}
\begin{ruledtabular}
\begin{tabular}{lcccc}
Particle  & Mode & Physics & \pt(\mt) range    & Ref.\\
          &      & Run     & GeV/$c$,GeV/$c^{2}$  & \\
\hline
$\pi^{0}$            & $\gamma\gamma$          & 5 & 0.5-20     & \cite{ppg063} \\
$\pi^{+}$, $\pi^{-}$ & ToF                     & 3 & 0.3-2.7    & \cite{ppg030} \\
$K^{+}$, $K^{-}$     & ToF                     & 3 & 0.4-1.9    & \cite{ppg030} \\
$K^{0}_{S}$          & $\pi^{0}\pi^{0}$        & 5 & 2-13.5     & \ref{tab:mesons} \\
$\eta$               & $\gamma\gamma$          & 3 & 2-12       & \cite{ppg055} \\
$\eta$               & $\gamma\gamma$          & 6 & 2-20       & \cite{ppg115} \\
$\eta$               & $\pi^{0}\pi^{+}\pi^{-}$ & 3 & 2.5-8.5    & \cite{ppg055} \\
$\omega$             & $e^{+}e^{-}$            & 5 & 0-4        & \ref{tab:omega} \\
$\omega$             & $\pi^{0}\pi^{+}\pi^{-}$ & 5 & 2-13.5     & \ref{tab:omega} \\
$\omega$             & $\pi^{0}\pi^{+}\pi^{-}$ & 3 & 2.5-10     & \cite{ppg064} \\
$\omega$             & $\pi^{0}\gamma$         & 5 & 2-12       & \ref{tab:omega}  \\
$\omega$             & $\pi^{0}\gamma$         & 3 & 2-7        & \cite{ppg064} \\
$\eta'$              & $\eta\pi^{+}\pi^{-}$    & 5 & 3-11.5     & \ref{tab:mesons}  \\
$\phi$               & $e^{+}e^{-}$            & 5 & 0-4        & \ref{tab:mesons} \\
$\phi$               & $K^{+}K^{-}$            & 5 & 1-8        & \ref{tab:mesons} \\
$J/\psi$             & $e^{+}e^{-}$            & 5 & 0-9        & \cite{ppg069} \\
$J/\psi$             & $e^{+}e^{-}$            & 6 & 0-9        & \cite{ppg097} \\
$\psi'$              & $e^{+}e^{-}$            & 6 & 0-7        & \cite{ppg104} \\
$p,\bar{p}$          & ToF	               & 3 & 0.6-3.7    & \cite{ppg030} \\
\end{tabular}
\end{ruledtabular}
\end{table}

\begin{figure}[htb]
\includegraphics[width=1.0\linewidth]{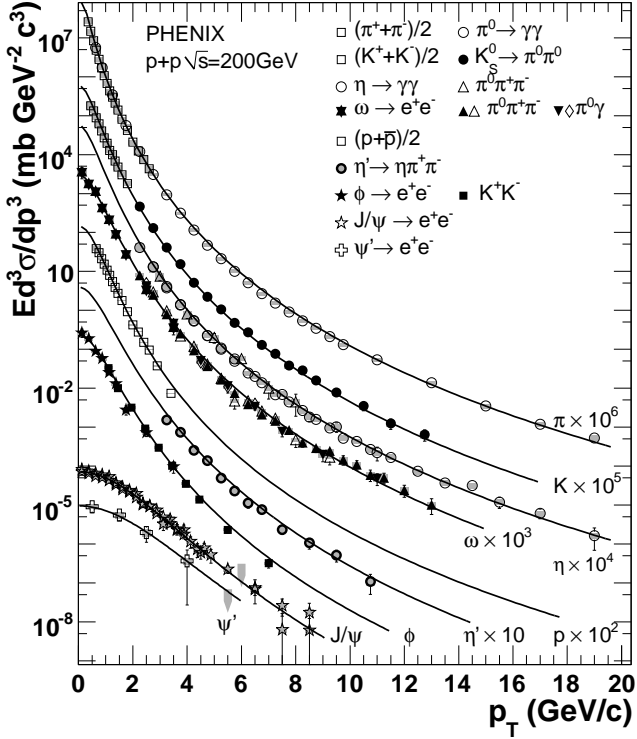}
\caption{Invariant differential cross sections of different particles 
measured in \pp collisions at \sqs=200~GeV in various decay modes. The 
spectra published in this paper are shown with closed symbols, previously 
published results are shown with open symbols. The curves are the fit 
results discussed in the text.
\label{fig:new}}
\end{figure}

Figure~\ref{fig:new} demonstrates a very good agreement between the 
new results and previously published data. The results presented in this 
paper greatly enhance the \pt range of the previously measured particles 
and add results for particles that have not been previously analyzed.

For each particle we considered all available measurements of the invariant 
momentum distributions together with their statistical and systematic 
uncertainties categorized as Type A, B, and C, as explained in 
section~\ref{sec:errors}.

For the analysis of the shape of the transverse momentum distributions the 
data for all particles of the same isospin multiplet were combined into one 
\pt spectrum to be fitted. All data for positively and negatively charged 
particles measured in the same analysis and in the same \pt bins were 
averaged. All data for neutral particles, measured via different decay 
channels, were added together. The notation $\pi$ is used to denote a 
combined spectrum of $\pi^0$ and $(\pi^{+}+\pi^{-})/2$, $K$ is used for a 
combined spectra of $K^0_{S}$ and $(K^{+}+K^{-})/2$, $p$ denotes 
$(p+\bar{p})/2$, and so forth. Independent measurements of the same 
particle performed using different data samples or different decay modes 
were also added together but not averaged. For data samples where the 
results were published as $dN/dp_T$ a conversion was made using 
Eq.~\ref{eq:cs}.

\subsection{Particle spectra fit distributions \label{sec:formula}}

It is widely known from experimental data that, as expected from pQCD 
calculations (e.g.~\cite{pqcd-tail}), a pure power law shape successfully 
describes the high \pt region of particle spectra: 
\begin{eqnarray} 
E\frac{d^{3}\sigma}{dp^{3}} = A p_{T}^{-\nu}
\label{eq:power}
\end{eqnarray} 
where the shape is determined by the power $\nu$ and $A$ is a 
normalization constant. However, the power law shape is seen to fail in 
the region below about \pt = 3--5~GeV/$c$ where the spectra exhibit a more 
exponential shape.

The exponential shape of the particle spectra at low \pt suggests a 
thermal interpretation in which the bulk of the produced particles are 
emitted by a system in thermal equilibrium with a Boltzmann-Gibbs 
statistical description of their spectra:
\begin{eqnarray} E\frac{d^{3}\sigma}{dp^{3}} = C_{b} e^{-E/T}
\label{eq:bg}
\end{eqnarray} 
where $C_{b}$ is a normalization factor and $E$ is the particle
energy. At midrapidity one can replace $E$ by $m_T =
(p_T^2 + m_0^2)^{1/2}$, where $m_0$ is the particle
rest mass. 

In recent years a variety of publications~\cite{Biyajima:2004ub,wilk_t,Tang:2008ud,Collaboration:2010xs,star_high_pt,star_kstar,tsallis_new} 
have used the Tsallis distribution~\cite{tsallis_0} to fit particle spectra.
The Tsallis distribution derives from a generalized form of the
Boltzmann-Gibbs entropy and is written as: 
\begin{eqnarray}
G_{q}(E) = C_{q}\left(1-(1-q)\frac{E}T\right)^{1/(1-q)}
\label{eq:tsallis}
\end{eqnarray} 
where $C_{q}$, $E$, and $T$ have similar meanings as in Eq.~\ref{eq:bg} 
and $q$ is the so-called nonextensivity parameter.  For values of $q\neq 
1$ the distribution exhibits a power law behavior with power $n = 
-1/(1-q)$. In order to associate the Tsallis distribution with a 
probability distribution, which describes the invariant particle spectra 
given by Eq.~\ref{eq:cs} and defined over $0 < E < \infty$, 
Eq.~\ref{eq:tsallis} must satisfy a normalization and energy conservation 
condition $\langle E \rangle < \infty$.  This limits the range of the 
parameter $q$ to $1<q<1\frac{1}{3}$.  The Tsallis distribution reduces to 
the Boltzmann-Gibbs distribution of Eq.~\ref{eq:bg} in the limit of $q 
\rightarrow 1$.

To put Eq.~\ref{eq:tsallis} into a form appropriate to fit particle spectra
we replace $E$ by $m_T = (p_T^2 + m_0^2)^{1/2}$ and use
the requirement of unit normalization to determine the
coefficient $C_{q}$ in Eq.~\ref{eq:tsallis} to be equal to: 
\begin{eqnarray} 
C_{q} & = & 
\frac{(2q-3)(q-2)}{T(T+m_0)-(q-1)(q-2)m_0^{2}} \nonumber \\
& & \times \frac{1}{\left(1-(1-q)\frac{m_0}T\right)^{1/(1-q)}}.
\end{eqnarray} 
Replacing the parameter $q$ with: 
\begin{eqnarray}
n & = &-\frac{1}{1-q} 
\label{eq:params}
\end{eqnarray}
The resulting formula used in the fitting procedure is given by: 
\begin{eqnarray}
E\frac{d^{3}\sigma}{dp^{3}} & = &
\frac{1}{2\pi}\frac{d\sigma}{dy}\frac{(n-1)(n-2)}{(nT+m_0(n-1))(nT+m_0)}
\nonumber \\
& & \times \left(\frac{nT+m_T}{nT+m_0}\right)^{-n}
\label{eq:levy}
\end{eqnarray} 
where \cs is the integrated cross section of the particle production
at midrapidity.

In the limit of $m_0 \rightarrow 0$ Eq.~\ref{eq:levy}  becomes:
\begin{eqnarray}
E\frac{d^{3}\sigma}{dp^{3}} = \frac{1}{2\pi}\frac{d\sigma}{dy}\frac{(n-1)(n-2)}{(nT)^{2}}
\left(1+\frac{m_T}{nT}\right)^{-n}.
\label{eq:hag}
\end{eqnarray}
This form is very similar to the QCD inspired expression suggested by Hagedorn in~\cite{Hagedorn:1983wk} written 
as a function of \mt instead of \pt. 

The condition that the shape of the \mt-spectra of different particles are 
the same regardless of their mass, is referred to as \mt-scaling. \mt 
scaling is known to provide a good description of the experimental data at 
low energy, where the spectral shapes are 
exponential~\cite{Guettler:1976fc, Bartke:1976zj}. Due to the explicit 
$m_0$ mass dependence in Eq.~\ref{eq:levy} the Tsallis distribution does 
not satisfy \mt scaling, except in the case $m_0 \rightarrow 0$ or $q 
\rightarrow 1$, in which case the limiting forms of Eqs.~\ref{eq:bg} or 
\ref{eq:hag} apply. Therefore the accuracy of fits to the Tsallis 
distribution and the validity of \mt scaling needs to be quantified with 
data.

The power law behavior at high \pt which appears in Eq.~\ref{eq:levy} is 
governed by the parameter $n$. The parameter $n$ can be related to the 
simple power law parameter $\nu$ that occurs in Eq.~\ref{eq:power} through 
the condition that both expressions have the same power-like slope at a 
given \pt.  From Eqs.~\ref{eq:levy}~and~\ref{eq:power} one can write:
\begin{eqnarray}
\frac{d {\rm ln}(nT+m_T)^{-n}}{d {\rm ln}(p_T)} = \frac{d {\rm ln}(p_{T}^{-\nu})}{d {\rm ln}(p_T)}\nonumber \\
n  = \frac{\nu m_{T}^{2}}{p_T^{2}-\nu T m_{T}}.
\label{eq:running}
\end{eqnarray} 
At high \pt ($p_{T}\gg m_{0},\nu T$) where one can neglect the difference between \mt and
\pt, $\nu$ and $n$ coincide.  In the \pt region where
most particle spectra are measured, $n$ is 15-25\%
larger than $\nu$.

The mean \mt of the Tsallis distribution in the form of Eq.~\ref{eq:levy} 
is calculated as:
\begin{eqnarray}
\langle m_T\rangle & = &\frac{2nT}{n-3}+\frac{(n-2)(n-1)}{(nT+m_0(n-1))(n-3)}m_0^{2} \nonumber \\
 & \approx &\frac{2nT}{n-3}+\frac{n-2}{n-3}m_0 .
\label{eq:mmt}
\end{eqnarray} 
The approximate relation requires $m_0 \gg T$. This condition is satisfied 
for all particles, except pions, for which $T$ and $m_0$ are about the 
same. Similarly, the mean \pt can be well approximated for all measured 
particles with a linear dependence:
\begin{eqnarray}
\langle p_T \rangle \approx \frac{2nT}{n-3} + f(n) m_0.
\label{eq:mpt}
\end{eqnarray} 
The first contribution is identical to that in Eq.~\ref{eq:mmt}
and $f(n)$ has only a weak dependence on $m_0$ which we neglect
in Eq.~\ref{eq:mpt}.

The Tsallis distribution is appealing to use to describe particle spectra 
because it provides a single functional form that can reproduce the full 
spectral shape with just two parameters that potentially have an 
underlying physical interpretation. Tsallis distributions have been used 
successfully to describe particle spectra in different collision systems 
and at different 
energies~\cite{star_high_pt,wilk-power,wilk-levy-cosmic,levy-ee,Biyajima:2004ub,wilk_t,Tang:2008ud,tsallis_new,Collaboration:2010xs}. 
Tsallis distributions also describe various physics phenomena beyond 
particle production and have been successfully applied in other fields of 
science, see~\cite{tsallis-thoughts,wilk-power,navarra-review,tsallis-web} 
and references therein.

As mentioned above, the Tsallis distribution was derived as the single 
particle distribution corresponding to a generalization of the 
Boltzmann-Gibbs entropy through the introduction of the non-extensivity 
parameter $q$~\cite{tsallis_0}. Whereas the Boltzmann-Gibbs distributions 
are found to apply to systems which exhibit an exponential relaxation in 
time to a stationary state characterized by exponentials in energy at 
thermal equilibrium, the generalized form is found to apply to systems 
which exhibit power laws in relaxation time and energy. These are systems 
which relax with a non-ergodic occupation of phase space as a consequence 
of the microscopic dynamics of the system.  Among other examples, this is 
characteristic of systems with long range interactions that fall off with 
distance with a power smaller than the dimensionality of the system.  It 
is an interesting question whether strongly interacting partonic matter 
might also exhibit power law relaxation. In fact, an analysis of the 
diffusion of a charmed quark in partonic matter produced in parton cascade 
calculations found that the parton densities were characterized by Tsallis 
distributions, rather than Boltzmann-Gibbs 
distributions~\cite{Walton:1999dy}.

The physical interpretation of the parameter $T$ in Eq.~\ref{eq:tsallis}, 
especially in \pp collisions, is not straightforward. One can expect that 
for larger systems, as produced in relativistic heavy ion collisions, it 
reflects the kinetic freeze-out temperature $\langle T_{{\rm kfo}} 
\rangle$ at which particle scattering ceases to modify the spectral 
shapes. It is shown below that the magnitudes of $\langle T \rangle$ found 
in this work are close to $\langle T_{{\rm kfo}} \rangle$ extracted in the 
blast-wave model approach~\cite{phenix-tfo,star_long} applied to \pp data. 
In high energy applications it has been shown~\cite{wilk_about_q} that the 
parameter $q$ of the Tsallis distribution of Eq.~\ref{eq:tsallis} can be 
related to the amount of temperature fluctuations in the system as:
\begin{eqnarray} 
q = 1+\frac{{{\rm Var}}(\frac{1}{T})}{\langle \frac{1}{T} \rangle^{2}}= 1 + \frac{1}{n}.
\label{eq:q}
\end{eqnarray}

\subsection{Fitting procedure \label{sec:fit_proc}}

In order to obtain a reliable estimate of the fit uncertainties, the 
experimental systematic uncertainties must be treated properly. The 
various types of systematic uncertainties have been taken into 
consideration as described here. The \pt independent systematic 
uncertainties of Type A have been combined in quadrature with the 
statistical errors and the \pt-independent systematic uncertainties of 
Type C was reduced by 9.7\% due to the trigger uncertainty, common to all 
analyzed particles. Residual uncertainties of Type C and of Type B must 
also be considered in the analysis. The Type B uncertainties by definition 
have an unknown \pt dependence. In order to estimate their effect, the 
particle spectra were varied and fit multiple times. For each fit the 
$y$-coordinate in each \pt bin was varied by the same amount according to 
the uncertainty of Type C, and by differing amounts according to the Type 
B uncertainties, in a manner similar to that explained in~\cite{ppg079}.

Variations of the $y$-coordinates were made independently for each fit 
with the amount of variation chosen randomly according to the 
\pt-dependent uncertainties for each particle and each sample. For the 
particle spectra consisting of multiple samples results of each fit to the 
entire spectrum were weighted with the probability of the fit estimated 
from the $\chi^{2}$ criteria. Such weighting emphasizes variations in 
which individual samples fluctuate toward each other rather than away from 
each other, which corresponds to the assumption that the different samples 
represent measurements of the same true momentum distribution.

As a result of the multiple fits, weighted distributions of the fit 
parameters were obtained.  The mean of the distribution was taken as the 
parameter value, the RMS width of the distribution was taken as the 
systematic uncertainty, and the statistical uncertainty was taken from the 
fit to the unmodified data. The number of fits was chosen such that the 
mean and the RMS did not change with increasing number of trials.

\subsection{Fit results \label{sec:fit_shape}}

\begin{figure}[htb]
\includegraphics[width=1.0\linewidth]{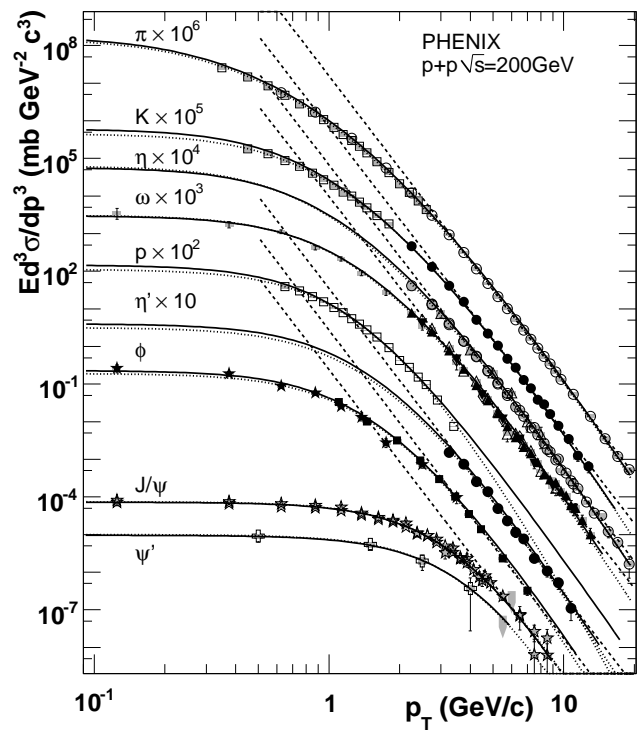}
\caption{The \pt spectra of various hadrons measured by PHENIX fitted 
to the power law (dashed lines) and Tsallis fit (solid lines). See text 
for more details. 
\label{fig:spec_pow_lv0}}
\end{figure}

The fits of Eq.~\ref{eq:levy} to the data are shown in 
Fig.~\ref{fig:spec_pow_lv0} with dotted lines. The results are given in 
Table~\ref{tab:free_fit}

\begin{table}[htb]
\caption{Parameters of the Tsallis fit with Eq.~\ref{eq:levy} with 
all parameters free to vary.   The uncertainties are statistical and 
systematic. Cross sections are in $\mu$b for $J/\psi$ and $\psi'$ and 
in mb for all other particles.\label{tab:free_fit}}
\begin{ruledtabular}
\begin{tabular}{lccc}
             &\cs (mb, $\mu$b)      & $T$ (MeV)                & $n=-1/(1-q)$          \\
\hline
$\pi$        & $43.5\pm2.0\pm1.9$   & $112.7\pm2.9\pm1.1$      & $9.57\pm0.11\pm0.03$  \\
$K$          & $4.0\pm0.1\pm0.5$    & $132.7\pm3.8\pm7.2$      & $10.04\pm0.16\pm0.27$ \\
$\eta$       & $5.1\pm1.1\pm3.9$    & $119\pm10\pm30$          & $9.68\pm0.18\pm0.49$  \\
$\omega$     & $4.3\pm0.3\pm0.4$    & $109.7\pm6.9\pm6.7$      & $9.78\pm0.24\pm0.18$  \\
$\eta'$      & $0.80\pm1.5\pm0.7$   & $141\pm107\pm61$         & $10.5\pm2.2\pm1.2$    \\
$\phi$       & $0.41\pm0.02\pm0.03$ & $139\pm16\pm15$          & $10.82\pm0.71\pm0.56$ \\
$J/\psi$     & $0.73\pm0.01\pm0.05$ & $149\pm56\pm82$          & $12.3\pm1.6\pm2.9$    \\
$\psi'$      & $0.13\pm0.03\pm0.02$ & $164\pm10^{3}\pm10^{2}$  & $14\pm12\pm6$         \\	
$p$          & $1.63\pm0.05\pm0.11$ & $107\pm13\pm12$          & $12.2\pm1.0\pm0.7$    \\
\end{tabular}
\end{ruledtabular}
\end{table}

The fit parameters $n$ and $T$ are strongly correlated. In some cases, the 
parameter $T$ can change by more than a factor of two, and still produce a 
good fit. Therefore additional information is needed to constrain the values 
of $n$ and $T$. For that purpose one can use a power law given by 
Eq.~\ref{eq:power} fitted to the same data. As discussed above the parameters 
$n$ and $\nu$ are related to each other through Eq.~\ref{eq:running}. However, 
it is found that the results of the power law fit depend on the fit range, but 
become stable when the fit range begins above \pt $\sim$ 3.5~GeV/$c$ for most 
particles, or above \pt $\sim$ 5.5~GeV/$c$ for heavier particles such as the 
$J/\psi$. The resulting power law fits are shown in 
Fig.~\ref{fig:spec_pow_lv0} as dashed lines that have been plotted down to 
\pt=0.5 GeV/$c$. Spectra without sufficient data above the fit range lower 
limit were not fitted. The results are given in Table~\ref{tab:fit_power}.

\begin{table}[htb]
\caption{Parameters of the power law fit with Eq.~\ref{eq:power}. The uncertainties are statistical 
and systematic. Units of $A$ are mb(GeV/$c)^{\nu+2}$. \label{tab:fit_power}}
\begin{ruledtabular}
\begin{tabular}{lccc}
             &$\nu$                   & $A$                  \\
\hline
$\pi$        &$8.174\pm0.035\pm0.049$ & $16.4\pm1.1\pm1.6$   \\
$K$          &$8.24\pm0.08\pm0.11$    & $8.8\pm0.9\pm1.6$    \\
$\eta$       &$8.169\pm0.037\pm0.054$ & $7.64\pm0.46\pm0.83$ \\
$\omega$     &$7.986\pm0.083\pm0.080$ & $9.5\pm1.3\pm1.4$    \\
$\eta'$      &$8.12\pm0.21\pm0.11$    & $3.6\pm1.2\pm0.8$    \\
$\phi$       &$8.20\pm0.36\pm0.15$    & $2.8\pm1.5\pm0.7$    \\
$J/\psi$     &$7.0\pm1.2\pm0.4$       & $0.03\pm0.03\pm0.02$ \\
\end{tabular}
\end{ruledtabular}
\end{table}

\begin{figure}[htb]
\includegraphics[width=1.0\linewidth]{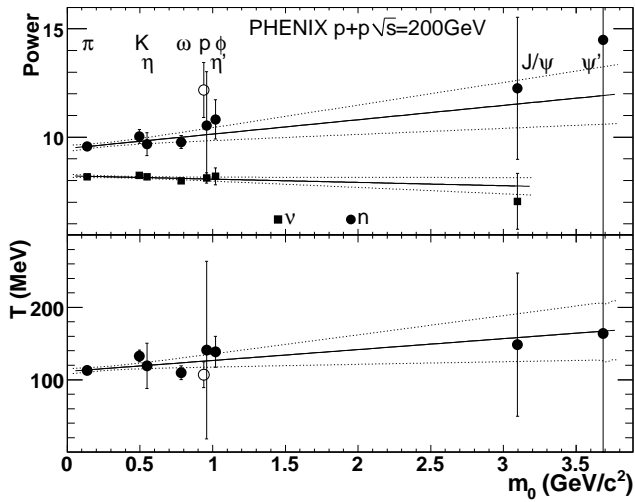}
\caption{Particle mass dependence of the fit parameters. Power law parameters 
$\nu$ and $n$ are plotted in the upper panel. Vertical bars denote the 
combined statistical and systematic uncertainties. The solid lines are linear 
fits. The dashed lines denote the fit uncertainty within which the linear fit 
can be inclined. The lower panel shows the same for the fit parameter $T$. The 
proton measurement (open circle) is not used in the fits.
\label{fig:param}}
\end{figure}

The parameters $\nu$ of the power law fits and the parameters $n$ and $T$ 
of the Tsallis fits are shown in Fig.~\ref{fig:param} as a function of the 
particle mass. The parameters have been fit to a linear function to 
establish if there is a mass dependence. The fits are shown in 
Fig.~\ref{fig:param} as solid lines with the uncertainties indicated by 
dashed lines. From Fig.~\ref{fig:param} it is evident that the parameters 
are consistent with no significant mass dependence. Therefore the 
parameters have also been fit with a constant value. The results for the 
linear and constant fits are summarized in Table.~\ref{tab:lin_fits}.

\begin{table}[htb]
\caption{Constant and linear fits to the power law and Tsallis 
fit parameters.  The last column (Prob.) gives the probability  
estimated by the $\chi^{2}/n.d.f.$ of the fit."
\label{tab:lin_fits}}
\begin{ruledtabular}
\begin{tabular}{llc}
           & Fit                                                & Prob.\\
$\nu$      & $~8.154\pm0.039$                                   & 0.75 \\
$\nu$      & $(8.22\pm0.07) - (0.15\pm0.14)m_{0}$[GeV/$c^{2}$]  & 0.79 \\
$n$        & $~9.656\pm0.097$                                   & 0.69 \\
$n$        & $(9.48\pm0.14) + (0.66\pm0.39)m_{0}$[GeV/$c^{2}$]  & 0.94 \\
$T$ (MeV)  & $~115.3\pm2.8$                                     & 0.43 \\
$T$ (MeV)  & $(111.5\pm4.0)     + (15\pm12)m_{0}$[GeV/$c^{2}$]  & 0.51 \\
\end{tabular}
\end{ruledtabular}
\end{table}

The fitted linear coefficients are consistent with zero within less than two 
standard deviation of the fit accuracy for all three parameters. At the same 
time the parameter $\nu$ is more accurately defined compared to the Tsallis 
fit parameter $n$. We can invoke Eq.~\ref{eq:running} to constrain the Tsallis 
fit using the parameter $\nu$. This requires to estimate the effective \pt 
which appears in Eq.~\ref{eq:running}. Using the mass independent terms of the 
fits listed in Table~\ref{tab:lin_fits} the effective \pt is about 7~GeV/$c$.

This value is large enough to allow to neglect the difference between \mt and 
\pt in Eq.~\ref{eq:running} for all particles, except the $J/\psi$ and 
$\psi'$. These two particles do not constrain the mass dependence of the 
Tsallis fit parameters due to their large fit uncertainties, as shown in 
Fig.~\ref{fig:param}.

Under the assumption that the parameter $\nu$ is the same for all particles, 
the mass dependence of the parameters $n$ and $T$ must either be present or 
absent together.  This can be checked by fixing the parameter $n$ to a 
constant value of $n =9.656$ (from Table~\ref{tab:lin_fits}) and fitting the 
data again. The mass dependent coefficient for the parameter $T$ that results 
in this case is somewhat different from zero compared to uncertainties. This 
is a clear contradiction to Eq.~\ref{eq:running} under the assumption of 
constant $\nu$, and therefore indicates that the parameters $n$ and $T$ have a 
mass dependence. However, this conclusion is at the limit of the accuracy of 
the currently available data.

For further analysis the parameter $n$ was fixed to have a linear dependence 
$n = 9.48+0.66m_{0}$~[GeV/$c^{2}$] (from Table~\ref{tab:lin_fits}) and the 
particle spectra were fit again. The results are given in 
Table~\ref{tab:fixed_n} and the fit to the mass dependence of $T$ is given in 
Table~\ref{tab:lin_fits_fixed}.

\begin{table}[htb]
\caption{Parameters of the Tsallis fit with Eq.~\ref{eq:levy} with parameter $n$ constrained to a 
fixed linear dependence on mass (for mesons). 
The uncertainties for \cs and $T$ are statistical and systematic, and are only systematic 
for $n$. Cross sections are in $\mu$b for $J/\psi$ and $\psi'$, and in mb for 
all other particles.\label{tab:fixed_n}}
\begin{ruledtabular}
\begin{tabular}{lccc}
             &\cs (mb, $\mu$b)         & $T$ (MeV)           & $n=-1/(1-q)$  \\
\hline
$\pi$        & $42.8\pm3.1\pm2.7$      & $112.6\pm2.1\pm2.8$ & $9.57\pm0.10$ \\
$K$          & $4.23\pm0.09\pm0.53$    & $125.4\pm0.9\pm5.3$ & $9.81\pm0.13$ \\
$\eta$       & $3.86\pm0.30\pm0.71$    & $124\pm2\pm12$      & $9.84\pm0.14$ \\
$\omega$     & $4.26\pm0.23\pm0.33$    & $115.5\pm2.1\pm6.8$ & $10.00\pm0.22$\\
$\eta'$      & $0.63\pm0.27\pm0.21$    & $123\pm17\pm18$     & $10.12\pm0.28$\\
$\phi$       & $0.427\pm0.019\pm0.023$ & $123.4\pm3.0\pm8.3$ & $10.16\pm0.31$\\
$J/\psi$     & $0.760\pm0.014\pm0.048$ & $148\pm8\pm35$      & $11.5\pm1.1$  \\
$\psi'$      & $0.132\pm0.029\pm0.020$ & $147\pm127\pm54$    & $11.9\pm1.3$  \\	
$p$          & $1.775\pm0.044\pm0.066$ & $58.8\pm1.8\pm6.1$  & $9.20\pm0.28$ \\
\end{tabular}
\end{ruledtabular}
\end{table}

Comparison of the results listed in 
Tables~\ref{tab:free_fit}~and~\ref{tab:fixed_n} reveals that the parameters of 
the fit did not change significantly within uncertainties, even for the $\eta$ 
and $\eta'$ mesons which are not measured at low \pt. In addition, with the 
$n$ parameter constrained the uncertainty on the parameter $T$ is reduced.

Since there is not yet a published PHENIX measurement of protons at high-\pt 
the parameter $\nu$ cannot be determined for the case of protons. Results 
published in~\cite{star_high_pt} suggest that the slope of the proton spectra 
at high \pt is the same as that for mesons. Using this assumption allows to 
extract the parameter $T$ for protons with the result listed in 
Table~\ref{tab:fixed_n}. The value of $T$ for protons differs from the values 
extracted for mesons.

\begin{table}[htb]
\caption{Constant and linear fits to the Tsallis parameter $T$ of 
mesons with fixed parameter $n$.  The last column (Prob.) gives the 
probability estimated by the $\chi^{2}/n.d.f.$ of the fit.
\label{tab:lin_fits_fixed}}
\begin{ruledtabular}
\begin{tabular}{llc}
           & Fit                                              & Prob.\\
$T$ (MeV)  &~$117.4\pm2.5$                                    & 0.64 \\
$T$ (MeV)  & $(112.6\pm3.8) + (11.8\pm7.0)m_{0}$[GeV/$c^{2}$] & 0.83 \\
\end{tabular}
\end{ruledtabular}
\end{table}

Using the linear dependence of the $T$ parameter 
$T=112.6+11.8m_{0}$[GeV/$c^{2}$] extracted from the fits to the Tsallis 
distribution with fixed linear dependence of the $n$ parameter (from 
Table~\ref{tab:fixed_n})  the spectra can be fit once again to obtain an 
improved normalization parameter. The resulting fits are shown in 
Fig.~\ref{fig:spec_pow_lv0} as the solid lines, and the results of the fit are 
given in Table~\ref{tab:fixed_n_T}.

\begin{table}[htb]
\caption{Parameters of the Tsallis fit with Eq.~\ref{eq:levy} with parameters $n$ and $T$ constrained 
to have a fixed linear dependence on mass (for mesons). 
The uncertainties for \cs are statistical and systematic, and are only systematic 
for $T$ and $n$. Cross sections are in $\mu$b for $J/\psi$ and $\psi'$, and in mb for 
all other particles.\label{tab:fixed_n_T}}
\begin{ruledtabular}
\begin{tabular}{lccc}
             &\cs (mb, $\mu$b)       & $T$ (MeV)    & $n=-1/(1-q)$      \\
\hline
$\pi$        & $40.5\pm0.3\pm5.8$      & $114.2\pm4.0$ & $9.57\pm0.10$  \\
$K$          & $4.71\pm0.06\pm0.48$    & $118.4\pm5.2$ & $9.81\pm0.13$  \\
$\eta$       & $4.46\pm0.05\pm0.97$    & $119.0\pm5.4$ & $9.84\pm0.14$  \\
$\omega$     & $3.64\pm0.07\pm0.77$    & $121.8\pm6.7$ & $10.00\pm0.22$ \\
$\eta'$      & $0.62\pm0.04\pm0.16$    & $123.8\pm7.7$ & $10.11\pm0.28$ \\
$\phi$       & $0.421\pm0.009\pm0.054$ & $124.5\pm8.1$ & $10.15\pm0.31$ \\
$J/\psi$     & $0.761\pm0.013\pm0.060$ & $149\pm22$    & $11.5\pm1.1$   \\
$\psi'$      & $0.133\pm0.024\pm0.019$ & $156\pm26$    & $11.9\pm1.3$   \\	
$p$          & $1.76\pm0.03\pm0.16$    & $58.8\pm6.4$  & $9.20\pm0.28$  \\
\end{tabular}
\end{ruledtabular}
\end{table}

The parameters $n$ and $T$, and their errors, are fixed to the values obtained 
from the fitted linear dependence of the parameters on particle mass, obtained 
from the fits of Tables~\ref{tab:lin_fits} and~\ref{tab:lin_fits_fixed}. The 
systematic error on the integrated yields reflects the variation of the $n$ 
and $T$ parameters within the errors. It also includes the uncertainty from 
the variation of the spectral shapes within errors of Types $B$ and $C$, as 
explained above.

\begin{figure*}[htb]
\includegraphics[width=0.7\linewidth]{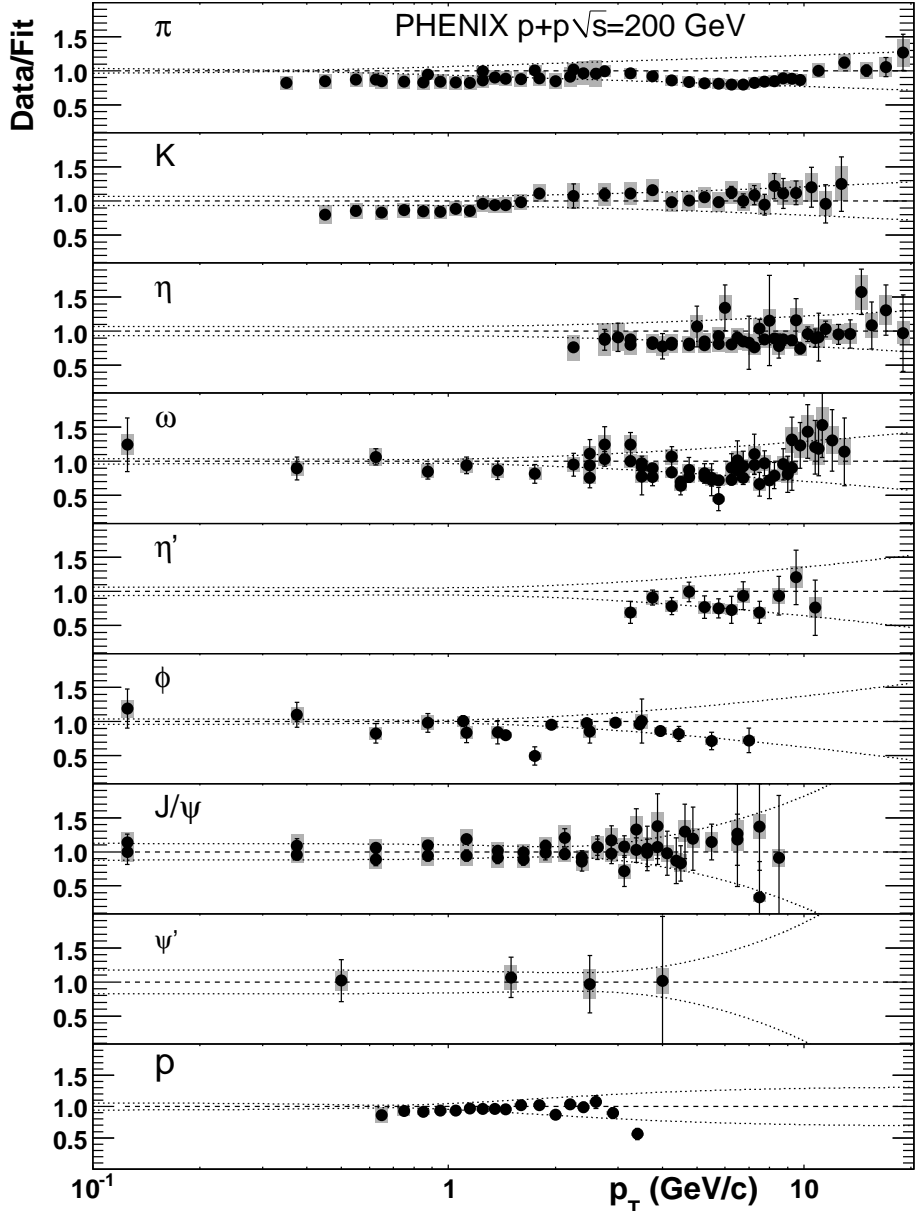}
\caption{Data to fit ratio for different particles used in the analysis. 
The systematic uncertainties are the combined uncertainties
of Type B and Type C, excluding the common  
9.7\%  trigger uncertainty.
\label{fig:data_to_fit}}
\end{figure*}

The fits accurately describe the data. To demonstrate the quality of the fits 
the data points have been divided by the fit value and the ratios plotted in 
Fig.~\ref{fig:data_to_fit}.

Grey error bars show the combined systematic uncertainty of Type B and C, with 
the Type B uncertainties dominating. The dashed lines show the fit uncertainty 
corridor. The RMS of the vertical spread of all points plotted in 
Fig.~\ref{fig:data_to_fit} is 0.17. If each point is normalized to the 
combined statistical and systematic error of the data point, the RMS of the 
same distribution is much larger with a value of 0.88, which indicates that 
the agreement between data and fit is well within errors.


\section{Discussion \label{sec:results}}

\subsection{Tsallis fit parameters}

The analysis of section~\ref{sec:ana_scaling} demonstrated the ability of the 
Tsallis distribution functional form to fit the full transverse momenta 
spectra for all different species produced in \pp collisions at \sqs = 200~GeV 
with only two parameters, $n=-1/(1-q)$ and $T$. Furthermore, the values of the 
two parameters extracted from the fits are approximately the same for all 
measured mesons.

On the other hand, the observation that the pure power law fit of 
Eq.~\ref{eq:power} to the spectra in the region of \pt$>$3.5~GeV/$c$ yields 
the same power $\nu=8.154\pm0.039$ for all particles with higher accuracy than 
the Tsallis fit, indicates that a weak mass dependence of the Tsallis 
parameters is to be expected. Assuming a weak mass dependence one gets 
$T=112.6\pm3.8 + (11.8\pm7.0)m_0 [{\rm GeV/c^2}]$~MeV and $n=9.48\pm0.14 + 
(0.66\pm0.39)m_0 [{\rm GeV/c^2}]$ that improves the description of the meson 
spectra with the Tsallis distribution. The parameters are listed in 
Tables~\ref{tab:fit_power}~and~\ref{tab:fixed_n_T} and plotted in 
Fig~\ref{fig:param}.

The ratios of the data points to the Tsallis parameterization using the global 
fit parameters $n$ and $T$ for all particles were shown in 
Fig.~\ref{fig:data_to_fit}. The figure represents nine different particles 
species measured over the range $0<p_{T}$~(GeV/$c)<20$ using six independent 
data samples and ten different analysis techniques. The parameterization is in 
good agreement with the experimental data. The average deviation of the points 
from one in all panels of the figure is 88\% of the combined uncertainty of 
the data and the fit.

The Tsallis distribution fit for the proton measurement yields a parameter 
$T=58.8 \pm 6.4$~MeV significantly lower than that for the mesons. Since the 
published PHENIX results for protons have limited \pt range this result was 
checked and confirmed using STAR measurements for protons and heavier 
baryons~\cite{star_high_pt,star_long,star_strange,star_strange_b,star_p_k_pi}. 
This result indicates significantly different Tsallis fit parameters between 
mesons and baryons.

The similarity of the measured parameters $T$ and $n$ for all studied mesons 
suggests a similar production mechanisms in \pp collisions at \sqs = 200~GeV. 
At the same time, the mechanism of baryon production must have different 
features. The interpretation of the $T$ parameter of the Tsallis fits is not 
straightforward. If interpreted as a temperature the values obtained are seen 
to be similar to average freeze-out temperatures $\langle T_{{\rm kfo}} 
\rangle$ extracted in the blast-wave model 
approach~\cite{phenix-tfo,star_long} applied to \pp data. As mentioned above, 
the parameter $n$ can be related to temperature fluctuations as 
$\sqrt{Var(1/T)}/\langle 1/T\rangle = 1/n$ in a thermal interpretation. 
Following this interpretation one can estimate the fluctuations of the inverse 
slope parameter $1/T$ to be of order of 0.3.

\subsection{\mt scaling}

As discussed in section~\ref{sec:formula}, \mt scaling can not be an exact 
scaling when particle spectra follow the Tsallis distribution with $q\neq 1$.  
However, \mt scaling might be found to be approximately true. The validity of 
\mt scaling can be studied quantitatively with the assistance of 
Eq.~\ref{eq:hag}, which gives the Tsallis distribution in the limit $m_0 
\rightarrow 0$ with a form explicitly satisfying \mt scaling.

\begin{figure}[htb]
\includegraphics[width=1.0\linewidth]{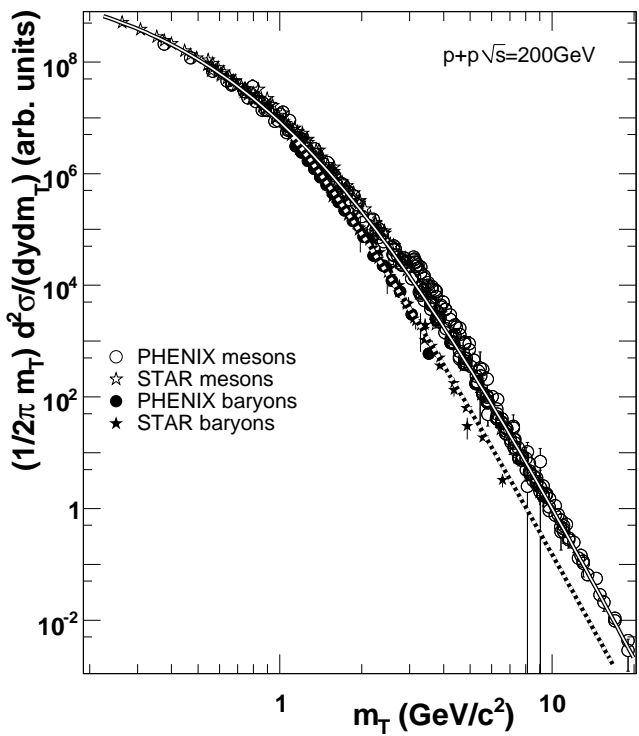}
\caption{Particle spectra plotted vs. \mt and arbitrarily normalized at 
\pt=10~GeV/$c$. Open symbols are mesons and full symbols are baryons measured 
by PHENIX (circles) and STAR (stars). The lines are the Hagedorn fits by 
Eq.~\ref{eq:hag} to mesons (solid) and baryons (dashed) with parameters $n$ 
and $T$ fixed to average values. Error bars are statistical and point-by-point 
systematic only.
\label{fig:mt}}
\end{figure}

Figure~\ref{fig:mt} shows the spectra for all particles plotted as a function 
of \mt and normalized at one single point on the X-axis. All normalized 
spectra are then fit simultaneously with Eq.~\ref{eq:hag} using fixed 
parameters taken from Tables~\ref{tab:lin_fits},\ref{tab:lin_fits_fixed} 
$n$=9.656 and $T=115.3$~MeV for mesons, and $T=58.8$~MeV for baryons.

The difference in the spectral shapes between mesons and baryons is apparent 
from the figure. It is due to the large difference in $\langle T \rangle$ 
between these particle groups. At the same time the spectra of both mesons and 
baryons separately are well described by the \mt scaling assumption.

To quantify this statement we restricted the analysis to the PHENIX meson 
measurements only. After optimization of the normalization point for the 
different particles the RMS of the data to fit ratio for all points shown in 
Fig~\ref{fig:mt} has a value of 0.25. This is to be compared to the analogous 
result of Fig.~\ref{fig:data_to_fit} for the Tsallis fit in \pt coordinate 
which gave an RMS of 0.17. This small increase supports the conclusion that at 
\sqs=200~GeV all meson spectra have very similar shape when plotted as a 
function of \mt, and thus obey \mt scaling.

\subsection{Integrated yields and \mpt}

Using the Tsallis functional form and 
Tables~\ref{tab:lin_fits},\ref{tab:lin_fits_fixed} one can derive information 
about \mmt and \mpt, based only on the particle mass and the baryon number. 
Determination of the integrated \cs requires experimental measurement of the 
particle production cross section in at least in a limited \pt range.

The results presented below were obtained independently for each particle 
species without averaging within the same isospin multiplet, unless such 
averaging was done by the experiment. Different measurements of the same 
particle were combined together. Published data from the STAR experiment and 
the references listed in Table~\ref{tab:results} were also analyzed. To 
compare PHENIX and STAR results, the spectra and the integrated yields 
published by STAR, in units of dN/dy, were multiplied by 30~mb, which is the 
value of the STAR minimum bias cross section in \pp collisions at \sqs = 
200~GeV, that includes the non-single diffractive part of \pp interactions 
(cf.~\cite{star_long}).

\begin{table*}[htb]
\caption{Cross sections in mb and \mpt in GeV/$c$ of different particles in 
\pp collisions at \sqs=200~GeV. PHENIX and STAR columns show the values 
obtained by fits to the experimental spectra with the Tsallis functional form 
as described in the text. One should state explicitly that these values do not 
supersede values given in the "Published" column by the experiments, in their 
publications listed in the last column, or elsewhere. An additional 9.7\% 
systematic uncertainty should be added to all $d\sigma/dy$ values listed in 
the column ``PHENIX'' and 12\% to the values in the column ``'STAR' to account 
for the trigger uncertainties. Values in the column ``Published'' are also 
given without these systematic uncertainties. The column ``S.M.'' is the 
prediction of the statistical model discussed in the text. The characteristic 
widths of the particle spectra are \mpt for all species except for $J/\psi$ 
and $\psi'$ for which the values given in the Table are $\langle 
p_{T}^{2}\rangle$. For $\psi'$ the integration is done in the \pt region below 
5~GeV/$c$. All errors are the combined statistical and systematic 
uncertainties.
\label{tab:results}}
\begin{ruledtabular}
\begin{tabular}{lccccccc}
                                      &\multicolumn{4}{c}{$d\sigma/dy$ (mb)} & \multicolumn{2}{c}{\mpt (GeV/$c$), $\langle p_{T}^{2}\rangle$ (GeV$^{2}/c^{2}$)} \\
Particle                              & PHENIX          & STAR            & Published       & S.M.  & Fit             & Published       & Ref.                  \\
\hline
$\pi^{0}$                             &$41.4\pm5.8$    &               &               & 46.9  &$0.377\pm0.012$&               &                       \\
$\pi^{+}$                             &$39.4\pm7.3$    &$43.8\pm3.3$   &$43.2\pm3.3$   & 42.1  &$0.379\pm0.012$&$0.348\pm0.018$& \cite{star_long}      \\
$\pi^{-}$                             &$38.6\pm7.2$    &$43.2\pm3.3$   &$42.6\pm3.3$   & 41.5  &$0.379\pm0.012$&$0.348\pm0.018$& \cite{star_long}      \\
$K^{+}$                               &$4.57\pm0.61$   &$4.72\pm0.39$  &$4.50\pm0.39$  & 4.57  &$0.567\pm0.017$&$0.517\pm0.030$& \cite{star_long}      \\
$K^{-}$                               &$4.20\pm0.51$   &$4.61\pm0.18$  &$4.35\pm0.39$  & 4.38  &$0.567\pm0.017$&$0.517\pm0.030$& \cite{star_long}      \\
$K^{0}_{S}$                           &$5.28\pm0.53$   &$4.26\pm0.15$  &$4.02\pm0.34$  & 4.40  &$0.569\pm0.017$&$0.605\pm0.025$& \cite{star_strange}   \\
$\eta$                                &$4.47\pm0.96$   & 	       &               & 4.93  &$0.595\pm0.018$& 	       &                       \\
$\rho$                                &		       &$6.55\pm0.37$  &$7.8\pm1.2$    & 5.58  &$0.714\pm0.019$&$0.616\pm0.062$& \cite{star_rho}       \\
$\omega$                              &$3.65\pm0.77$   & 	       &$4.20\pm0.47$  & 5.03  &$0.718\pm0.022$&$0.664\pm0.039$& this work             \\
$\eta'$                               &$0.62\pm0.17$   & 	       &               & 0.365 &$0.808\pm0.026$& 	       &                       \\
$(K^{*+}+K^{*-})/2$                   &		       &$1.46\pm0.10$  &               & 1.57  &$0.774\pm0.022$& 	       &                       \\
$(K^{*0}+\bar{K}^{*0})/2$             &		       &$1.525\pm0.091$&$1.52\pm0.19$  & 1.55  &$0.776\pm0.022$&$0.81\pm0.14$  & \cite{star_kstar}     \\
$\phi$                                &$0.421\pm0.055$ & 	       &$0.432\pm0.035$& 0.339 &$0.839\pm0.027$&$0.752\pm0.043$& this work             \\
$\phi$                                &		       &$0.525\pm0.018$&$0.540\pm0.086$& 0.339 &$0.839\pm0.025$&$0.820\pm0.051$& \cite{star_phi}       \\
$J/\psi$ ($\times10^{3}$)             &$0.759\pm0.053$ & 	       &$0.746\pm0.089$&       &$4.464\pm0.606$&$4.60\pm0.19$  & \cite{ppg069}         \\
$\psi'$  ($\times10^{3}$)             &$0.133\pm0.031$ & 	       &$0.126\pm0.034$&       &$4.807\pm0.443$&$4.7\pm1.3$    & \cite{ppg104}         \\
\hline
$p$                                   &                &$4.06\pm0.23$  &$4.14\pm0.30$  & 4.47  &$0.648\pm0.019$&$0.661\pm0.022$& \cite{star_long}      \\
$\bar{p}$                             &                &$3.28\pm0.23$  &$3.39\pm0.36$  & 3.59  &$0.648\pm0.019$&$0.661\pm0.022$& \cite{star_long}      \\
$\Lambda$                             &		       &$1.33\pm0.13$  &$1.31\pm0.12$  & 1.30  &$0.742\pm0.023$&$0.775\pm0.040$& \cite{star_strange}   \\
$\bar{\Lambda}$                       &		       &$1.20\pm0.12$  &$1.19\pm0.11$  & 1.11  &$0.742\pm0.023$&$0.763\pm0.040$& \cite{star_strange}   \\
$\Xi^{-}$                             &		       &$0.094\pm0.020$&$0.078\pm0.028$& 0.092 &$0.850\pm0.030$&$0.924\pm0.054$& \cite{star_strange}   \\
$\bar{\Xi}^{+}$                       &		       &$0.091\pm0.019$&$0.087\pm0.031$& 0.082 &$0.850\pm0.030$&$0.881\pm0.051$& \cite{star_strange}   \\
$\Sigma^{*+}+\Sigma^{*-}$             &		       &$0.358\pm0.026$&$0.321\pm0.044$& 0.308 &$0.882\pm0.032$&$1.020\pm0.073$& \cite{star_strange_b} \\
$\bar{\Sigma}^{*+}+\bar{\Sigma}^{*-}$ &		       &$0.310\pm0.025$&$0.267\pm0.038$& 0.260 &$0.882\pm0.032$&$1.010\pm0.061$& \cite{star_strange_b} \\
$\bar{\Lambda}^{*}+\Lambda^{*}$       &		       &$0.127\pm0.013$&$0.104\pm0.017$& 0.168 &$0.955\pm0.038$&$1.08\pm0.10$  & \cite{star_strange_b} \\
$\Omega^{-}+\bar{\Omega}^{+}$ ($\times10^{3}$) &       &$11.5\pm4.6$   &$10.2\pm5.7$   & 17.1  &$1.035\pm0.046$&$1.08\pm0.30$  & \cite{star_strange}   \\
\end{tabular}
\end{ruledtabular}
\end{table*}

The particle spectra published by the STAR experiment were fit to the Tsallis 
functional form given by Eq.~\ref{eq:levy} with the parameters 
$n=9.48+0.66m_0[{\rm GeV/c^2}]$ and $T =112.6+11.8m_0[{\rm GeV/c^2}]$ taken 
from the global fit to the PHENIX data. The same parameters determined 
independently for the STAR data give consistent results for mesons. For 
baryons the STAR data showed a dependence of the parameter $T$ on the mass of 
the particle, however the fit uncertainties were too large to make a definite 
statement. The value of $T$ averaged over all baryon measurements made by STAR 
agrees with the PHENIX result for the proton measurement. Calculation of \cs 
for $p$ and $\bar{p}$ measured by PHENIX was not done because the spectra are 
feed-down corrected and the extrapolation to low \pt requires additional 
evaluation of the systematic uncertainties.

\begin{figure}[htb]
\includegraphics[width=1.0\linewidth]{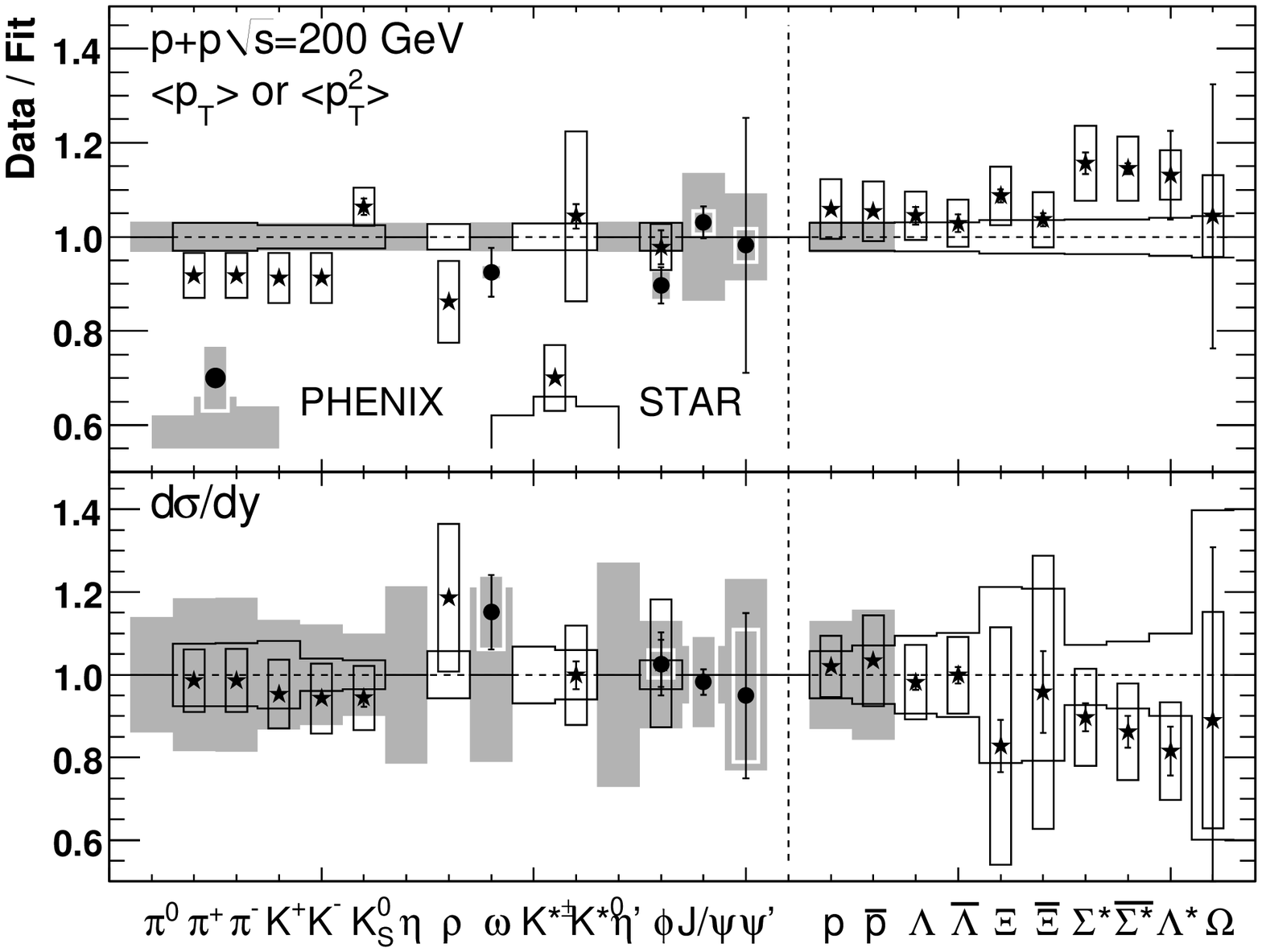}
\caption{Comparison of the integrated parameters of the particle spectra. The 
upper panel shows the ratio of the published result divided by the result of 
constrained Tsallis fit for \mpt ($\langle p_T^{2} \rangle$ for the $J/\psi$ 
and the same for $\psi'$ integrated in the \pt range below 5~GeV/$c$). The 
lower panel shows the ratio for \cs. The statistical and systematic 
uncertainties shown at each point are from the published data only. The band 
around $y=1$ shows the uncertainty of the values extracted from the Tsallis 
function. The trigger efficiency scale uncertainty of 9.7\% for PHENIX and 
12\% for STAR is not plotted. There are no points outside the plot boundaries. 
Vertical dashed line separates mesons and baryons.
\label{fig:res_ratios}}
\end{figure}

Figure~\ref{fig:res_ratios} shows a comparison of the experimentally measured 
integrated spectral characteristics to the results obtained using the Tsallis 
fits. The ratio of the measured characteristic width to the width calculated 
from the Tsallis fit is shown in the upper panel. For most particles the width 
is taken to be \mpt, but for the $J/\psi$ and $\psi'$ the comparison is done 
for $\langle p_T^{2} \rangle$ because this is the parameter published in the 
corresponding articles. Statistical and systematic uncertainties of the 
published results are shown at each data point and the uncertainties of the 
Tsallis fit values are shown by the band around $y=1$.

For all mesons the agreement between the published values and the values from 
the Tsallis fit analysis is consistent with the published uncertainties. This 
demonstrates the accuracy to which the Tsallis functional form describes the 
experimental spectral shapes.

Eq.~\ref{eq:mpt} suggests that the mass dependence of the \mpt should be 
approximately linear. A fit to the average mean momentum of all mesons 
extracted from the Tsallis distribution fits as a function of their mass gives 
$\langle p_{T}\rangle=(0.319\pm0.007)[$GeV/$c]+(0.491\pm0.009)m_{0}$.  A fit 
to the published data directly gives a similar consistent result of $\langle 
p_{T}\rangle=(0.284\pm0.015)[$GeV/$c]+(0.506\pm0.033)m_{0}$. For baryons the 
agreement with the linear fit is reasonable based on the data published by the 
STAR experiment.

In the original work of R.~Hagedorn~\cite{Hagedorn:1983wk} a nearly linear 
dependence of the \mpt was derived based on the assumption of Boltzmann-Gibbs 
statistics to describe the particle spectra at low \pt. The difference between 
mesons and baryons would follow from the bosonic and fermionic nature of these 
particles. However, quantitatively the values of particle \mpt and the 
magnitude of the meson-to-baryon difference are not the same as would follow 
from the mechanisms discussed in~\cite{Hagedorn:1983wk}.

The lower panel of Fig.~\ref{fig:res_ratios} shows the ratio of the integrated 
yields published by the experiment to the integrated yields extracted from the 
Tsallis function fits. The common uncertainties on all integrated yields of 
9.7\% for PHENIX and 12\% for STAR are not included. Most of the ratios equal 
1 within uncertainties. From Fig.~\ref{fig:res_ratios} and 
Table~\ref{tab:results} one may conclude that the constrained Tsallis fit 
reproduces the measured integrated cross section with high accuracy for all 
identified particles in \pp collisions at \sqs = 200~GeV. This gives 
justification to use the constrained Tsallis fit results to obtain \cs for 
particles which have only been measured in a limited \pt range, such as 
$\pi^0$, $\eta$, and $\eta'$ mesons. The resulting \cs for such particles
 are also given in Table~\ref{tab:results}.

It should be noted explicitly that the \cs and \mpt values given in 
Table~\ref{tab:results} determined using the Tsallis parameterization do not 
supersede, or presume to be more accurate than the corresponding values 
published by the experiments in the original papers. They are given to 
validate the method.  In those cases where no values have been published the 
Tsallis fit result values in the table represent a best attempt to obtain the 
cross section or \mpt based on the validity of the Tsallis fit distribution.

\subsection{Statistical model calculation}

\begin{figure}[htb]
\includegraphics[width=1.0\linewidth]{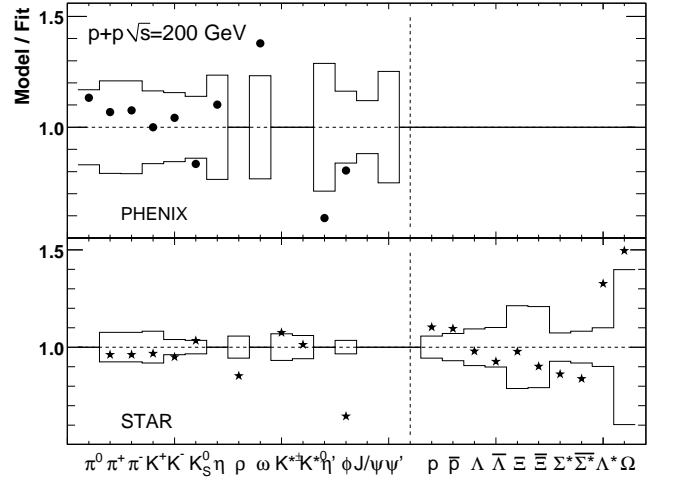}
\caption{Ratio of integrated yields predicted by the statistical model~\cite{we} to those 
of the constrained Tsallis fits for various particles. Results for fits to PHENIX data are shown in the upper panel and STAR data 
in the lower panel. The band reflects the uncertainty of the Tsallis fit results and includes the
trigger uncertainty of 9.7\% for PHENIX. The lower panel has smaller uncertainties because the 
model prediction was based on published STAR data given in Table~\ref{tab:results}.
There are no points outside the plot boundaries. Vertical dashed line separates mesons 
and baryons.\label{fig:res_ratios_sm}}
\end{figure}

Figure~\ref{fig:res_ratios_sm} shows the ratio of the constrained Tsallis fit 
results for the integrated particle yields to the predicted yields from a 
statistical model (SM) calculation~\cite{sm}. The data-to-model ratio for 
PHENIX data is shown in the upper panel, and for STAR data in the lower panel. 
The statistical model calculation parameters were chosen to reproduce the 
integrated yields published by the STAR 
experiment~\cite{star_long,star_strange,star_strange_b,star_phi,star_rho,star_kstar}, 
which may explain the larger discrepancies in the comparison to the PHENIX 
results.

Although statistical models are not commonly used to describe \pp data the 
agreement of the statistical model calculation with the STAR results was found 
to be accurate for most particles except for the $\rho$, $\phi$, and 
$\Lambda^{*}$~\cite{we}. Leaving aside baryons, for which the calculations of 
the \cs requires additional assumptions, as explained above, the Tsallis fit 
also has difficulty to reproduce the result for the $\rho$ meson as shown in 
Fig.~\ref{fig:res_ratios}. This can be explained by the large systematic 
uncertainty of the published value~\cite{star_rho}.

For the PHENIX data the SM calculations agree with the production rates for 
most mesons because the Tsallis fit results of the PHENIX and STAR data agree. 
The production rates of $\pi^0$, $\eta$, $\omega$, $\eta'$, and $\phi$ were 
not measured by STAR and so were not used in the determination of the SM 
parameters. Among them the predicted yields of $\pi^0$ and $\eta$ mesons are 
in very good agreement with the PHENIX data. The prediction of the SM for the 
$\omega$, $\eta'$, and $\phi$ yields are less accurate with ratios just 
outside of errors.

\section{Summary \label{sec:summary}}

A systematic study of neutral meson production in \pp collisions at 
\sqs=200~GeV has been performed by the PHENIX experiment at RHIC with results 
presented in this paper. New measurements by PHENIX of $K_{S}^{0}$, $\omega$, 
$\phi$, and $\eta'$ meson production have been presented.

The measurement of the $K_{S}^0$ invariant differential cross section via the 
$\pi^0\pi^0$ decay channel in the momentum range $2<p_{T}$~(GeV/$c)<13.5$ 
extends previously published $K^\pm$ measurements~\cite{ppg030}.

We present the first measurement of the $\phi$ invariant differential cross 
section in the $K^+K^-$ decay mode using several different techniques. The 
combined spectrum reaches the upper \pt limit of 8~GeV/$c$.

This work also presents the first measurement of the invariant differential 
cross section of $\eta'$ production measured via the $\eta\pi^+\pi^-$ decay 
mode with results that cover the range $3<p_{T}$~(GeV/$c)<11$.

Measurements of $\omega$ meson production in non-leptonic decay channels 
extends the \pt coverage of the previous PHENIX $\omega$ 
measurement~\cite{ppg064}, obtained with a smaller data sample, to 
13.5~GeV/$c$.

First measurements of the $\omega$ and $\phi$ in the $e^{+}e^{-}$ decay 
channel extend the \pt coverage for these two particles down to zero momentum 
and allow a direct calculation of the integrated yields and mean transverse 
momenta with results: $d\sigma^{\omega}/dy=4.20\pm0.33^{\rm stat}\pm0.52^{\rm 
syst}$~mb and $d\sigma^{\phi}/dy=0.432\pm0.031^{\rm stat}\pm0.051^{\rm 
syst}$~mb;  and $\langle p_{T}^{\omega}\rangle=0.664\pm0.037^{\rm 
stat}\pm0.012^{\rm syst}$~GeV/$c$ and $\langle 
p_{T}^{\phi}\rangle=0.752\pm0.032^{\rm stat}\pm0.014^{\rm syst}$~GeV/$c$.

All measured results were found to be consistent between the different decay 
modes and analysis techniques, as well as with previously published data. The 
results are shown in Figs.~\ref{fig:new_tranc} and~\ref{fig:new}, and 
the measured cross sections are tabulated in the Appendix.

The invariant differential cross sections for all measured hadrons produced in 
\pp collisions at \sqs=200~GeV presented in this work as well as in previous 
PHENIX publications, were shown to be described well over the entire momentum 
range by the Tsallis distribution functional form with only two parameters, 
$T$ and $n$, characterizing the low- and high-\pt regions, respectively. 
Furthermore, the values of the two parameters extracted from the fits are 
approximately the same for all measured mesons with a weak mass dependence: $ 
T =112.6 \pm 3.8 + (11.8\pm7.0)m_0 [{\rm GeV/c^2}]$~MeV; and $n=9.48 \pm 0.14 
+ (0.66\pm0.39)m_0 [{\rm GeV/c^2}]$.

The meson spectral shapes have very similar forms when plotted as a function 
of \mt and hence follow \mt-scaling well at \sqs=200~GeV. On the other hand, 
the proton spectra are described with a significantly lower parameter value of 
$T=58.8\pm 6.4$~MeV and do not follow the \mt-scaling form observed for 
mesons.

The ability to successfully describe all particle spectra in \pp collisions at 
\sqs = 200~GeV with a common functional form allows one to calculate the 
invariant differential cross section for any particle. This allows the 
absolute integrated yield to be derived from any experimental measurement of 
the hadron spectrum, even with limited \pt range. The values of \cs and \mpt 
are tabulated in Table~\ref{tab:results} for hadrons measured by PHENIX, as 
well as those measured by the STAR experiment using the set of values of 
Tables~\ref{tab:lin_fits}~and~\ref{tab:lin_fits_fixed}. For all measured 
mesons the average transverse momentum of the particle depends linearly on the 
mass $m_0$ and can be parameterized with the relation $\langle 
p_{T}\rangle=(0.319\pm0.007)[$GeV/$c]+(0.491\pm0.009)m_{0}$

The predictions of statistical model calculations based on data published by 
the STAR experiment~\cite{we} were shown to be in good agreement with the 
integrated yields calculated from the Tsallis distribution fits for most 
particles. Some deviations are seen for the $\omega$, the $\eta'$, and the 
$\phi$ mesons.


\section*{ACKNOWLEDGMENTS}   

We thank the staff of the Collider-Accelerator and Physics 
Departments at Brookhaven National Laboratory and the staff of the 
other PHENIX participating institutions for their vital 
contributions.  We also thank F. Becattini of the University of 
Florence for providing us the statistical model predictions and 
G. Wilk of SINS, Warsaw for useful discussions about applications 
of the Tsallis distribution.
We acknowledge support from the Office of Nuclear Physics in the
Office of Science of the Department of Energy,
the National Science Foundation, 
a sponsored research grant from Renaissance Technologies LLC, 
Abilene Christian University Research Council, 
Research Foundation of SUNY, 
and Dean of the College of Arts and Sciences, Vanderbilt University 
(U.S.A),
Ministry of Education, Culture, Sports, Science, and Technology
and the Japan Society for the Promotion of Science (Japan),
Conselho Nacional de Desenvolvimento Cient\'{\i}fico e
Tecnol{\'o}gico and Funda\c c{\~a}o de Amparo {\`a} Pesquisa do
Estado de S{\~a}o Paulo (Brazil),
Natural Science Foundation of China (People's Republic of China),
Ministry of Education, Youth and Sports (Czech Republic),
Centre National de la Recherche Scientifique, Commissariat
{\`a} l'{\'E}nergie Atomique, and Institut National de Physique
Nucl{\'e}aire et de Physique des Particules (France),
Ministry of Industry, Science and Tekhnologies,
Bundesministerium f\"ur Bildung und Forschung, Deutscher
Akademischer Austausch Dienst, and Alexander von Humboldt Stiftung (Germany),
Hungarian National Science Fund, OTKA (Hungary), 
Department of Atomic Energy (India), 
Israel Science Foundation (Israel), 
National Research Foundation (Korea),
Ministry of Education and Science, Russia Academy of Sciences,
Federal Agency of Atomic Energy (Russia),
VR and the Wallenberg Foundation (Sweden), 
the U.S. Civilian Research and Development Foundation for the
Independent States of the Former Soviet Union, 
the US-Hungarian Fulbright Foundation for Educational Exchange,
and the US-Israel Binational Science Foundation.




\section*{APPENDIX} 

\label{sec:appendix_a}

Tables of the measured invariant differential cross section $\frac{1}{2\pi
p_{T}}\frac{d^2\sigma}{dydp_{T}}$.

\begingroup \squeezetable

\begin{table*}
\caption{The invariant differential cross section $\frac{1}{2\pi 
p_{T}}\frac{d^2\sigma}{dydp_{T}}$ of $\omega$ meson production measured in the 
indicated decay channel.  Notations are:  $V$ is the differential cross section, 
$A$, $B$, and $C$ are the three types of errors described in the text.
\label{tab:omega}}
\begin{ruledtabular}
\begin{tabular}{ccccccc}
meson & decay &
\pt     & $V$ & $A$ & $B$ & $C$ \\
 & channel &
GeV/$c$ &  \multicolumn{4}{c}{mb/(GeV/$c)^{2}$} \\
\hline
& &   0.125 &   3.5                 &   1.1                &   0.4                &   0.2                \\	
& &   0.375 &   1.76                &   0.33               &   0.19               &  $8. \times10^{-2}$  \\	
& &   0.625 &   1.12                &   0.12               &   0.12               &  $5. \times10^{-2}$  \\	
& &   0.875 &   0.425               &  $5.6\times10^{-2}$  &  $4.6\times10^{-2}$  &  $1.9\times10^{-2}$  \\	
$\omega$ & $e^{+}e^{-}$ &
      1.125 &   0.213               &  $2.8\times10^{-2}$  &  $2.3\times10^{-2}$  &  $2. \times10^{-3}$  \\	
& &   1.375 &  $9.0 \times10^{-2}$  &  $1.4\times10^{-2}$  &  $1. \times10^{-2}$  &  $4. \times10^{-3}$  \\	
& &   1.75  &  $2.73\times10^{-2}$  &  $4.6\times10^{-3}$  &  $2.9\times10^{-3}$  &  $1.2\times10^{-3}$  \\	
& &   2.5   &  $3.34\times10^{-3}$  &  $6.3\times10^{-4}$  &  $3.5\times10^{-4}$  &  $1.5\times10^{-4}$  \\	
& &   3.5   &  $3.6 \times10^{-4}$  &  $1.2\times10^{-4}$  &  $4. \times10^{-5}$  &  $2. \times10^{-5}$  \\	
\hline
& &   2.25  &  $8.0 \times10^{-3}$  &  $1.4\times10^{-3}$  &  $8. \times10^{-4}$  &  $2.6\times10^{-4}$  \\	
& &   2.75  &  $2.50\times10^{-3}$  &  $2.5\times10^{-4}$  &  $2.4\times10^{-4}$  &  $8. \times10^{-5}$  \\	
& &   3.25  &  $7.89\times10^{-4}$  &  $5.8\times10^{-5}$  &  $7.4\times10^{-5}$  &  $2.6\times10^{-5}$  \\	
& &   3.75  &  $2.56\times10^{-4}$  &  $1.7\times10^{-5}$  &  $2.3\times10^{-5}$  &  $8. \times10^{-6}$  \\	
& &   4.25  &  $9.41\times10^{-5}$  &  $5.9\times10^{-6}$  &  $8.0\times10^{-6}$  &  $3.1\times10^{-6}$  \\	
& &   4.75  &  $3.69\times10^{-5}$  &  $2.5\times10^{-6}$  &  $3.5\times10^{-6}$  &  $1.2\times10^{-6}$  \\	
& &   5.25  &  $1.68\times10^{-5}$  &  $1.3\times10^{-6}$  &  $1.5\times10^{-6}$  &  $5. \times10^{-7}$  \\	
& &   5.75  &  $7.57\times10^{-6}$  &  $7.1\times10^{-7}$  &  $7.1\times10^{-7}$  &  $2.5\times10^{-7}$  \\	
& &   6.25  &  $3.89\times10^{-6}$  &  $4.1\times10^{-7}$  &  $3.7\times10^{-7}$  &  $1.3\times10^{-7}$  \\	
& &   6.75  &  $2.13\times10^{-6}$  &  $2.8\times10^{-7}$  &  $2.2\times10^{-7}$  &  $7. \times10^{-8}$  \\	
$\omega$ & $\pi^{0}\pi^{+}\pi^{-}$ &
      7.25  &  $1.45\times10^{-6}$  &  $2.2\times10^{-7}$  &  $1.6\times10^{-7}$  &  $5. \times10^{-8}$  \\	
& &   7.75  &  $8.5 \times10^{-7}$  &  $1.6\times10^{-7}$  &  $1.0\times10^{-7}$  &  $3. \times10^{-8}$  \\	
& &   8.25  &  $4.03\times10^{-7}$  &  $9.8\times10^{-8}$  &  $4.6\times10^{-8}$  &  $1.3\times10^{-8}$  \\	
& &   8.75  &  $2.93\times10^{-7}$  &  $7.0\times10^{-8}$  &  $3.6\times10^{-8}$  &  $1.0\times10^{-8}$  \\	
& &   9.25  &  $2.48\times10^{-7}$  &  $6.2\times10^{-8}$  &  $3.5\times10^{-8}$  &  $8. \times10^{-9}$  \\	
& &   9.75  &  $1.49\times10^{-7}$  &  $4.0\times10^{-8}$  &  $2.2\times10^{-8}$  &  $5. \times10^{-9}$  \\	
& &   10.25 &  $1.09\times10^{-7}$  &  $3.0\times10^{-8}$  &  $1.7\times10^{-8}$  &  $4. \times10^{-9}$  \\	
& &   10.75 &  $6.0 \times10^{-8}$  &  $1.9\times10^{-8}$  &  $1. \times10^{-8}$  &  $2. \times10^{-9}$  \\	
& &   11.25 &  $5.1 \times10^{-8}$  &  $1.6\times10^{-8}$  &  $8. \times10^{-9}$  &  $2. \times10^{-9}$  \\	
& &   12.   &  $2.41\times10^{-8}$  &  $8.3\times10^{-9}$  &  $3.6\times10^{-9}$  &  $8. \times10^{-10}$ \\	
& &   13.   &  $1.02\times10^{-8}$  &  $4.4\times10^{-9}$  &  $1.7\times10^{-9}$  &  $3. \times10^{-10}$ \\	
\hline
& &   2.5   &  $4.15\times10^{-3}$  &  $5.0\times10^{-4}$  &  $5.5\times10^{-4}$  &  $2.3\times10^{-4}$  \\	
& &   3.5   &  $4.54\times10^{-4}$  &  $3.2\times10^{-5}$  &  $4.9\times10^{-5}$  &  $2.5\times10^{-5}$  \\	
& &   4.5   &  $5.07\times10^{-5}$  &  $4.8\times10^{-6}$  &  $5.2\times10^{-6}$  &  $2.8\times10^{-6}$  \\	
& &   5.5   &  $1.14\times10^{-5}$  &  $1.8\times10^{-6}$  &  $1.1\times10^{-6}$  &  $6. \times10^{-7}$  \\	
$\omega$ & $\pi^{0}\gamma$ &
      6.5   &  $3.33\times10^{-6}$  &  $4.5\times10^{-7}$  &  $3.8\times10^{-7}$  &  $1.8\times10^{-7}$  \\	
& &   7.5   &  $7.7 \times10^{-7}$  &  $2.0\times10^{-7}$  &  $1.0\times10^{-7}$  &  $4. \times10^{-8}$  \\	
& &   9.    &  $1.94\times10^{-7}$  &  $6.4\times10^{-8}$  &  $2.9\times10^{-8}$  &  $1.1\times10^{-8}$  \\	
& &   11.   &  $4.8 \times10^{-8}$  &  $1.6\times10^{-8}$  &  $1.0\times10^{-8}$  &  $3. \times10^{-9}$  \\	
\end{tabular}
\end{ruledtabular}
\end{table*}

\begin{table*}
\caption{The invariant differential cross section $\frac{1}{2\pi 
p_{T}}\frac{d^2\sigma}{dydp_{T}}$ of $K_{S}^{0}$, $\eta'$, and $\phi$
meson production measured in the indicated 
decay channel.  Notations are:  $V$ is the differential cross section,
$A$, $B$, and $C$ are the three types of errors described in the text.
\label{tab:mesons}}
\begin{ruledtabular}
\begin{tabular}{ccccccc}
meson & decay &
\pt     & $V$ & $A$ & $B$ & $C$ \\
 & channel &
GeV/$c$ &  \multicolumn{4}{c}{mb/(GeV/$c)^{2}$} \\
\hline
& &   2.25  &  $4.66\times10^{-3}$  &  $7.7\times10^{-4}$  &  $7.4\times10^{-4}$  &  $3.0\times10^{-4}$  \\	
& &   2.75  &  $1.30\times10^{-3}$  &  $1.1\times10^{-4}$  &  $1.8\times10^{-4}$  &  $8.0\times10^{-5}$  \\	
& &   3.25  &  $4.14\times10^{-4}$  &  $2.5\times10^{-5}$  &  $5.5\times10^{-5}$  &  $2.6\times10^{-5}$  \\	
& &   3.75  &  $1.54\times10^{-4}$  &  $8.0\times10^{-6}$  &  $2.0\times10^{-5}$  &  $1.0\times10^{-5}$  \\
& &   4.25  &  $5.09\times10^{-5}$  &  $2.8\times10^{-6}$  &  $6.6\times10^{-6}$  &  $3.2\times10^{-6}$  \\	
& &   4.75  &  $2.22\times10^{-5}$  &  $1.2\times10^{-6}$  &  $2.9\times10^{-6}$  &  $1.4\times10^{-6}$  \\	
& &   5.25  &  $1.06\times10^{-5}$  &  $6.0\times10^{-7}$  &  $1.4\times10^{-6}$  &  $7.0\times10^{-7}$  \\	
& &   5.75  &  $4.74\times10^{-6}$  &  $3.3\times10^{-7}$  &  $6.1\times10^{-7}$  &  $3.0\times10^{-7}$  \\	
& &   6.25  &  $2.74\times10^{-6}$  &  $2.2\times10^{-7}$  &  $3.6\times10^{-7}$  &  $1.7\times10^{-7}$  \\	
$K_{S}^{0}$ & $\pi^{0}\pi^{0}$ &
      6.75  &  $1.30\times10^{-6}$  &  $1.3\times10^{-7}$  &  $1.7\times10^{-7}$  &  $8.0\times10^{-8}$  \\	
& &   7.25  &  $7.70\times10^{-7}$  &  $1.0\times10^{-7}$  &  $1.0\times10^{-7}$  &  $5.0\times10^{-8}$  \\	
& &   7.75  &  $3.82\times10^{-7}$  &  $6.0\times10^{-8}$  &  $5.3\times10^{-8}$  &  $2.4\times10^{-8}$  \\	
& &   8.25  &  $2.88\times10^{-7}$  &  $4.4\times10^{-8}$  &  $4.1\times10^{-8}$  &  $1.8\times10^{-8}$  \\	
& &   8.75  &  $1.59\times10^{-7}$  &  $3.1\times10^{-8}$  &  $2.3\times10^{-8}$  &  $1.0\times10^{-8}$  \\	
& &   9.50  &  $7.80\times10^{-8}$  &  $1.2\times10^{-8}$  &  $1.1\times10^{-8}$  &  $4.9\times10^{-9}$  \\	
& &   10.5  &  $3.49\times10^{-8}$  &  $8.5\times10^{-9}$  &  $5.6\times10^{-9}$  &  $2.2\times10^{-9}$  \\	
& &   11.5  &  $1.25\times10^{-8}$  &  $3.7\times10^{-9}$  &  $2.2\times10^{-9}$  &  $8.0\times10^{-10}$ \\	
& &   12.75 &  $6.50\times10^{-9}$  &  $2.1\times10^{-9}$  &  $1.3\times10^{-9}$  &  $4.1\times10^{-10}$ \\	
\hline
& &   3.25  &  $1.53\times10^{-4}$  &  $3.6\times10^{-5}$  &  $1.6\times10^{-5}$  &  $7. \times10^{-6}$  \\	
& &   3.75  &  $7.38\times10^{-5}$  &  $8.9\times10^{-6}$  &  $7.3\times10^{-6}$  &  $3.3\times10^{-6}$  \\	
& &   4.25  &  $2.55\times10^{-5}$  &  $4.1\times10^{-6}$  &  $2.5\times10^{-6}$  &  $1.1\times10^{-6}$  \\	
& &   4.75  &  $1.39\times10^{-5}$  &  $2.0\times10^{-6}$  &  $1.3\times10^{-6}$  &  $6. \times10^{-7}$  \\	
& &   5.25  &  $4.9 \times10^{-6}$  &  $1.0\times10^{-6}$  &  $5. \times10^{-7}$  &  $2. \times10^{-7}$  \\	
& &   5.75  &  $2.32\times10^{-6}$  &  $4.3\times10^{-7}$  &  $2.3\times10^{-7}$  &  $1.0\times10^{-7}$  \\	
$\eta\prime$ & $\eta\pi^{+}\pi^{-}$ & 
      6.25  &  $1.13\times10^{-6}$  &  $3.1\times10^{-7}$  &  $1.1\times10^{-7}$  &  $5. \times10^{-8}$  \\	
& &   6.75  &  $7.7 \times10^{-7}$  &  $1.7\times10^{-7}$  &  $8. \times10^{-8}$  &  $3. \times10^{-8}$  \\	
& &   7.5   &  $2.33\times10^{-7}$  &  $5.4\times10^{-8}$  &  $2.7\times10^{-8}$  &  $1.0\times10^{-8}$  \\	
& &   8.5   &  $1.07\times10^{-7}$  &  $3.3\times10^{-8}$  &  $1.2\times10^{-8}$  &  $5. \times10^{-9}$  \\	
& &   9.5   &  $5.2 \times10^{-8}$  &  $1.7\times10^{-8}$  &  $6. \times10^{-9}$  &  $2. \times10^{-9}$  \\	
& &   10.75 &  $1.09\times10^{-8}$  &  $5.8\times10^{-9}$  &  $2.1\times10^{-9}$  &  $4.9\times10^{-10}$ \\	
\hline
& &   0.125 &   0.264               &  $6.3\times10^{-2}$  &  $2.6\times10^{-2}$  &  $1.1\times10^{-2}$  \\	
& &   0.375 &   0.188               &  $3.1\times10^{-2}$  &  $1.8\times10^{-2}$  &  $8. \times10^{-3}$  \\	
& &   0.625 &  $8.9 \times10^{-2}$  &  $1.5\times10^{-2}$  &  $9. \times10^{-3}$  &  $4. \times10^{-3}$  \\	
& &   0.875 &  $5.83\times10^{-2}$  &  $8.2\times10^{-3}$  &  $5.8\times10^{-3}$  &  $2.5\times10^{-3}$  \\	
& &   1.125 &  $2.57\times10^{-2}$  &  $4.3\times10^{-3}$  &  $2.5\times10^{-3}$  &  $1.2\times10^{-3}$  \\	
$\eta'$ & $\eta\pi^{+}\pi^{-}$ & 
      1.375 &  $1.31\times10^{-2}$  &  $2.7\times10^{-3}$  &  $1.3\times10^{-3}$  &  $6. \times10^{-4}$  \\	
& &   1.75  &  $2.79\times10^{-3}$  &  $7.5\times10^{-4}$  &  $2.8\times10^{-4}$  &  $1.3\times10^{-4}$  \\	
& &   2.5   &  $7.2 \times10^{-4}$  &  $1.5\times10^{-4}$  &  $7. \times10^{-5}$  &  $3. \times10^{-5}$  \\	
& &   3.5   &  $9.7 \times10^{-5}$  &  $3.1\times10^{-5}$  &  $1.0\times10^{-5}$  &  $4. \times10^{-6}$  \\	
\hline
& &   1.1   &  $3.32\times10^{-2}$  &  $2.6\times10^{-3}$  &  $2.6\times10^{-3}$  &  $4. \times10^{-4}$  \\	
& &   1.45  &  $1.01\times10^{-2}$  &  $7. \times10^{-4}$  &  $5. \times10^{-4}$  &  $1. \times10^{-4}$  \\	
& &   1.95  &  $3.16\times10^{-3}$  &  $1.9\times10^{-4}$  &  $1.7\times10^{-4}$  &  $4. \times10^{-5}$  \\	
& &   2.45  &  $9.28\times10^{-4}$  &  $5.6\times10^{-5}$  &  $5.3\times10^{-5}$  &  $1.1\times10^{-5}$  \\	
& &   2.95  &  $2.99\times10^{-4}$  &  $1.9\times10^{-5}$  &  $1.8\times10^{-5}$  &  $4. \times10^{-6}$  \\	
$\phi$ & $K^{+}K^{-}$ &
      3.45  &  $1.02\times10^{-4}$  &  $6. \times10^{-6}$  &  $6. \times10^{-6}$  &  $1. \times10^{-6}$  \\	
& &   3.95  &  $3.49\times10^{-5}$  &  $2.6\times10^{-6}$  &  $2.3\times10^{-6}$  &  $4. \times10^{-7}$  \\	
& &   4.45  &  $1.38\times10^{-5}$  &  $1.8\times10^{-6}$  &  $9. \times10^{-7}$  &  $2. \times10^{-7}$  \\	
& &   5.5   &  $2.31\times10^{-6}$  &  $4.1\times10^{-7}$  &  $1.6\times10^{-7}$  &  $3. \times10^{-8}$  \\	
& &   7.    &  $3.21\times10^{-7}$  &  $7.9\times10^{-8}$  &  $2.4\times10^{-8}$  &  $4. \times10^{-9}$  \\	
\end{tabular}
\end{ruledtabular}
\end{table*}

\endgroup

\clearpage



\end{document}